\documentclass[twocolumn,aps,prl,showpacs,superscriptaddress,preprintnumbers,floatfix,nofootinbib]{revtex4-1}
\usepackage{multirow}
\usepackage{hhline}
\usepackage{bbm}
\usepackage{epsfig}
\usepackage{graphics}  
\usepackage{graphicx}  
\usepackage{dcolumn}   
\usepackage{bm}        
\usepackage{amssymb}   
\usepackage{amsmath}   
\usepackage{amsfonts}
\usepackage[percent]{overpic}
\usepackage{natbib}
\usepackage{hyperref}
\usepackage[usenames]{color}
\usepackage{CJK}

\newcommand{\sqrtsnn}{\mbox{$\sqrt{s^{}_{\mathrm{NN}}}$}}
\newcommand{\pT} {p_{\mathrm{T}}}

\newcommand{\ruru}{$^{96}$Ru+$^{96}$Ru}
\newcommand{\zrzr}{$^{96}$Zr+$^{96}$Zr}

\usepackage{tikz,xcolor,hyperref}

\definecolor{lime}{HTML}{A6CE39}
\DeclareRobustCommand{\orcidicon}{
	\begin{tikzpicture}
	\draw[lime, fill=lime] (0,0) 
	circle [radius=0.16] 
	node[white] {{\fontfamily{qag}\selectfont \tiny ID}};
	\draw[white, fill=white] (-0.0625,0.095) 
	circle [radius=0.007];
	\end{tikzpicture}
	\hspace{-2mm}
}
\foreach \x in {A, ..., Z}{%
	\expandafter\xdef\csname orcid\x\endcsname{\noexpand\href{https://orcid.org/\csname orcidauthor\x\endcsname}{\noexpand\orcidicon}}
}
\foreach \x in {A, ..., Z}{%
	\expandafter\xdef\csname orcid\x\endcsname{\noexpand\href{https://orcid.org/\csname orcidauthor\x\endcsname}{\noexpand\orcidicon}}
}

\begin{document}

\title{Disentangling nuclear structure through multiparticle azimuthal correlations in high-energy isobar collisions}

\newcommand{\moe}{Key Laboratory of Nuclear Physics and Ion-beam Application (MOE), and Institute of Modern Physics, Fudan
University, Shanghai 200433, China}
\newcommand{\fudan}{Shanghai Research Center for Theoretical Nuclear Physics, NSFC and Fudan University, Shanghai 200438, China}
\newcommand{\sbu}{Department of Chemistry, Stony Brook University, Stony Brook, NY 11794, USA}
\newcommand{\bnl}{Physics Department, Brookhaven National Laboratory, Upton, NY 11976, USA}
\newcommand{\ecnu}{School of Physics, East China Normal University, Shanghai 200062, China}

\author{Zaining Wang\orcidA{}}\affiliation{\moe}\affiliation{\sbu}
\author{Jinhui Chen\orcidB{}}
\email{chenjinhui@fudan.edu.cn}\affiliation{\moe}\affiliation{\fudan}
\author{Jiangyong Jia\orcidC{}}\email{jiangyong.jia@stonybrook.edu}\affiliation{\sbu}\affiliation{\bnl}
\author{Yu-Gang Ma\orcidD{}}\email{mayugang@fudan.edu.cn}\affiliation{\moe}\affiliation{\fudan}\affiliation{\ecnu}
\author{Chunjian Zhang\orcidE{}}\email{chunjianzhang@fudan.edu.cn}\affiliation{\moe}\affiliation{\fudan}\affiliation{\sbu}

\begin{abstract}
Event-by-event fluctuations in the amplitudes of flow harmonics offer a novel approach to probing the initial-state characteristics in heavy-ion collisions. In this study, we conduct a systematic investigation of correlations among various flow harmonics utilizing multiparticle cumulants in $^{96}$Ru+$^{96}$Ru and $^{96}$Zr+$^{96}$Zr collisions at $\sqrtsnn =$ 200 GeV within the framework of a multiphase transport model. Correlated nuclear density distributions specific to the isobar systems are incorporated to evaluate the sensitivity of selected observables to variations in nuclear deformation and neutron skin thickness. The analysis reveals that multiparticle azimuthal correlations are responsive to these nuclear structure features, predominantly in the most central collision events. Furthermore, the examined correlations exhibit shallow dependence on the assumed shear viscosity values. These findings provide a quantitative evaluation of the extent to which multiparticle flow observables can discern nuclear structure effects in isobar collisions and offer valuable guidance for future detailed dynamical investigations and experimental measurements.
\end{abstract}
\maketitle
\label{sec:intro}

\textit{Introduction.---}
\label{intro}
Anisotropic flow serves as a key observable for studying the properties of quark-gluon plasma (QGP), the hot and dense matter created in ultra-relativistic heavy-ion collision~\cite{Ollitrault:1992bk,Teaney:2000cw}. Within hydrodynamics descriptions, the initial pressure gradients convert spatial anisotropies in the collision geometry into azimuthal modulations of final-state particle distributions. These modulations are conventionally quantified via a Fourier decomposition of the azimuthal particle distribution,
\begin{equation}\begin{split}
\label{eq:1}
\mathrm{d}N/\mathrm{d} \phi \propto 1+2 \sum_{n=1}^{\infty} v_{n} \cos n\left(\phi-\mathrm{\Psi}_{n}\right),
\end{split}
\end{equation}
where $N$ is the number of particles produced, $\phi$ is the azimuthal angle of the particle, $v_n$ and $\mathrm{\Psi_n}$ represent the magnitude and the event-plane angle of the $n^{th}$ flow harmonic~\cite{Voloshin:2008dg,Gale:2013da,Heinz:2013th,Romatschke:2017ejr,Alqahtani:2025wan}, respectively. The anisotropic flow coefficients $v_n$ exhibit a good approximate proportionality to the initial-state eccentricities $\epsilon_n$, i.e., $v_{n} \simeq  \kappa_{n} \varepsilon_{n}$ (n=2,3), where the response coefficient $\kappa_n$ encodes the medium’s transport properties, particularly the shear viscosity to entropy density ratio $\eta/s$~\cite{Teaney:2012ke,Deng:2024} and the equation of state~\cite{PhysRevLett.114.202301,Voloshin:2008dg}. Deviations from this simple linear mapping arise from event-by-event fluctuations and nonlinear mode couplings, motivating the study of higher-order correlations and multiparticle observables.

Over the past decades, extensive measurements of the $``$flow paradigm" of produced particles at both the Relativistic Heavy Ion Collider (RHIC)~\cite{STAR:2015mki,STAR:2020gcl,STAR:2022gki,STAR:2022pfn} and the Large Hadron Collider (LHC)~\cite{ATLAS:2019peb,ALICE:2016kpq,CMS:2017glf} have provided a wealth of data, including single flow harmonics, event-by-event flow fluctuations, and correlations among different flow harmonics. These measurements have substantially restricted theoretical models of the initial conditions and have enhanced our quantitative understanding of the transport properties of the QGP.

The spatial distributions of nucleons, governed by the structure of the colliding atomic nuclei, constitute a primary source of the initial-state geometric anisotropies and determine the transverse size of the QGP overlap region. In quantum many-body systems, the nuclear structure itself is a fundamental property of the atomic nuclei, emerging from the collective and correlated dynamics of nucleons~\cite{Jia:2025wey,Ke:2025tyv,Giacalone:2023hwk}. In recent years, increasing attention has been paid to the role of nuclear structure effects in relativistic heavy-ion collisions, motivated by the realization that ground-state nuclear properties can influence the initial conditions of the collision system. Such effects have been explored using both soft and hard probes within modern hydrodynamic and hybrid model frameworks~\cite{Zhao:2022grq,Kharzeev:2022hqz,PhysRevC.106.014906,Shou:2014eya,PhysRevLett.127.242301,Nie:2022gbg,Jia:2021tzt,Xu:2021uar,Jia:2021qyu,Liu:2022kvz,PhysRevC.106.034913,Zhao:2022uhl,Jia:2022qrq,Giacalone:2019pca,Jia:2021wbq,Bally:2021qys,Nijs:2021kvn,Jia:2021oyt,Chen:2024zwk,Xu:2024bdh,Wang:2024vjf,Zhao:2024lpc,Xu:2021qjw,Zhao:2022mce,Fortier:2023xxy,Fortier:2024yxs,Zhang:2024bcb,Jia:2024xvl,Magdy:2022cvt,MaYG:2023,XiBS:2025,GG:2024,SB:2024,JJ:2025}. These high-energy studies demonstrate that nuclear deformations and neutron skin effects can leave measurable imprints on selected observables across energy scales.

Experimental measurements of multiparticle azimuthal correlation in $^{238}$U+$^{238}$U and $^{197}$Au+$^{197}$Au collisions at RHIC~\cite{STAR:2024wgy,Zhang:2022sgk}, as well as in $^{129}$Xe+$^{129}$Xe and $^{208}$Pb+$^{208}$Pb collisions at the LHC~\cite{ATLAS:2022dov,ALICE:2021gxt,ALICE:2024nqd}, have reported patterns consistent with expectations from nuclear quadrupole and, in some cases, triaxial deformations. These observations support the notion that intrinsic nuclear geometry can influence flow-related observables in relativistic heavy-ion collisions. Similarly, isobar collisions performed at RHIC involving nuclei with the same mass number, $^{96}$Ru+$^{96}$Ru and $^{96}$Zr+$^{96}$Zr, have provided direct evidence of nuclear structure difference, including quadrupole and octupole deformations as well as neutron skin thickness. These differences manifest themselves in the ratios of final-state bulk observables at a given centrality, where systematic uncertainties from hadronic evolution are expected to be largely canceled out~\cite{Zhang:2022fou,Jia:2022ozr,Xu:2022ikx,Zhang:2024ake}. Despite these advances, the behavior and sensitivity of multiparticle anisotropic-flow–related observables in isobar collisions have not yet been systematically assessed.

In this paper, we investigate $^{96}$Ru+$^{96}$Ru and $^{96}$Zr+$^{96}$Zr isobar collisions at $\sqrtsnn =$ 200 GeV using a multi-phase transport (AMPT) framework. This analysis focuses on the correlated long-range mixed-harmonic fluctuations as supplementary observables, including asymmetric cumulants, symmetric cumulants, nonlinear coupling coefficients, and the correlations between flow harmonics and mean transverse momentum. These observables are examined as potential probes of initial-state geometry and fluctuations, with particular attention to their achievable sensitivity to differences in nuclear structure parameters within the limitations of the model and available statistics. We also assess the dependence of isobar ratios on the choice of shear viscosity within the explored AMPT configurations.

This paper is organized as follows. Section II describes the general setup of the AMPT model and details our analysis methodology. Section III presents the main results and their implications. Section IV summarizes the findings and discusses their implications for future experimental studies.

\textit{Model setup and observables.---}\label{subsec:hydro}
\label{sec:1}
The AMPT transport model~\cite{Lin:2004en} is a hybrid framework with four main phases: (i) fluctuating initial conditions from the HIJING model~\cite{PhysRevD.44.3501}, (ii) elastic parton cascade simulated by the ZPC model~\cite{Zhang:1997ej}, (iii) quark coalescence model for hadronization~\cite{He:2017tla,Shao:2020sqr}, and (iv) hadronic re-scattering based on the ART model~\cite{Li:1995pra}. The spatial distribution of nucleons in the initial conditions is sampled using a deformed Woods-Saxon (WS) density profile of the form
\begin{equation}\begin{split}
\label{eq:1}
\rho(r, \theta)=\frac{\rho_{0}}{1+e^{\left[r-R_0(1+\beta_2Y_{2,0}(\theta)+\beta_3Y_{3,0}(\theta)) / a_{0}\right]}},
\end{split}
\end{equation}
where $\rho_0$ denotes the nucleon density at the center of the nucleus, and $a_0$ is the surface diffuseness parameter, also known as the skin depth. The half-density nucleus radius is $R_0=1.2A^{1/3}$ where $A$ is the mass number. The $\beta_n$ values can be estimated from the measured reduced electric transition probability $B(En)\uparrow$ via the standard formula $\beta_n=\left(4 \pi / 3 Ze R_0^n\right) \sqrt{B(En) \uparrow}$ with $Z$ the charge number of the nucleus~\cite{Moller:2015fba}. $\beta_2$ and $\beta_3$ represent the quadrupole and octupole deformation parameters respectively and account for the deformation in the nuclear geometry. The deviation from the spherical axially symmetric components is parameterized by spherical harmonics expressed as, $Y_{2,0}(\theta)=\frac{1}{4} \sqrt{\frac{5}{\pi}}\left(3 \cos ^{2} \theta-1\right)$ and $Y_{3,0}(\theta)=\frac{1}{4} \sqrt{\frac{7}{\pi}}\left(5 \cos ^{3} \theta-3 \cos \theta\right)$, where $\theta$ is the polar angle with respect to the symmetry axis of the nucleus.
\begin{table}[!h]
\caption{Nuclear structure parameters appearing in Eq.~\eqref{eq:1} and employed in the simulations of \ruru{} and \zrzr{} collisions at $\sqrtsnn =$ 200 GeV. Case1 and Case5 represent, respectively, the full parameterizations of $^{96}$Ru and $^{96}$Zr.}
\label{tab:1}
\centering
\begin{tabular}{l|cccc}\hline\hline
   &\; $R_0$ (fm)\; & \;$a_{0}$ (fm)\;  & $\beta_{2}$ & $\beta_{3}$  \\\hline
Case1 $^{96}$Ru & 5.09  & 0.46   & 0.162 & 0  \\
Case2          & 5.09  & 0.46   & 0.06  & 0  \\
Case3          & 5.09  & 0.46   & 0.06 & 0.20  \\
Case4          & 5.09  & 0.52   & 0.06 & 0.20  \\
Case5 $^{96}$Zr & 5.02  & 0.52   & 0.06 & 0.20  \\\hline\hline
$\vphantom{\displaystyle\frac{A_A}{A_A}}$ Ratios & \multicolumn{4}{c}{$\frac{\mathrm{\textstyle\small Case1}}{\mathrm{ \textstyle\small Case2}}$\;,\; $\frac{\mathrm{\textstyle\small Case1}}{\mathrm{ \textstyle\small Case3}}$\;,\;$\frac{\mathrm{\textstyle\small Case1}}{\mathrm{ \textstyle\small Case4}}$\;,\;$\frac{\mathrm{\textstyle\small Case1}}{\mathrm{ \textstyle\small Case5}}$}\\\hline\hline
\end{tabular}
\end{table}

Following our previous work~\cite{Jia:2021oyt,Zhang:2022fou,Zhang:2021kxj,Jia:2022qgl}, we perform simulations of isobar collisions using the AMPT model with version $v2.26t5$ in string-melting mode. For each collision configuration, we generate high computing challenging minimum-bias events with a minimum of 300 million, in order to reduce statistical uncertainties in multiparticle correlation observables. The partonic cross section is set to 3 $mb$ and is evaluated through $\sigma=9/2 \pi \alpha_{s}^{2}/\mu^{2}$, where $\alpha_s$ represents the QCD coupling constant, and $\mu$ denotes the screening mass. This particular AMPT configuration has demonstrated remarkable success in reproducing experimental measurements of multiparticle azimuthal correlation across both RHIC and the LHC energies~\cite{Lin:2001zk,Lin:2021mdn,Shen:2021pds,Zhang:2021ygs,Zhang:2022fum,Shao:2022eyd,Zhang:2025yyd}. The five distinct configurations of the nuclear parameters $\beta_2$, $\beta_3$, $R_0$, and $a_0$ are systematically varied in Tab.~\ref{tab:1}. It allows us to isolate the influences of the nuclear structure parameters step by step. Specifically, Case1/Case2 indicates the effect of $\beta_2$, Case1/Case3 isolates the effects of $\beta_2$ and $\beta_3$, Case1/Case4 probes the effects of $\beta_2$, $\beta_3$ and $a_0$, and Case1/Case5 examines the effects of $\beta_2$, $\beta_3$, $a_0$ and $R_0$.

In addition, we also study the dependence of $\eta/s$ on the flow harmonics. For a system of two massless quark flavors at initial temperature $T$, the $\eta/s$ can be estimated using the following pocket formula~\cite{Xu:2011fi}
\begin{eqnarray}\label{eq:2}
\frac{\eta}{s}\approx\frac{3 \pi}{40 \alpha_{s}^{2}} \frac{1}{\left(9+\frac{\mu^{2}}{T^{2}}\right) \ln \left(\frac{18+\mu^{2} / T^{2}}{\mu^{2} / T^{2}}\right)-18}.
\end{eqnarray}
Assuming an initial temperature of $T=0.38$~GeV, this would correspond to a $\eta/s$ value of 0.232 at an early time. The partonic cross-section dependence manifests in the transport coefficients as follows: a 1.5 $m$b cross-section corresponds to larger $\eta/s$ values, while the cases of 6.0 $m$b and 10.0 $m$b produce smaller $\eta/s$ values~\cite{Xu:2011fi}, as quantified in Tab.~\ref{tab:2}. We emphasize that the absolute shear viscosity values are less significant than their relative differences, which enable a robust examination of the isobar ratio systematics.
\begin{table}[h!]
\caption{The parameter settings in the AMPT model, where the $\eta/s$ value is estimated from Eq.~\eqref{eq:2} assuming an initial temperature of $T=0.38$~GeV.}
\label{tab:2}
\centering
\begin{tabular}{lccc}
\hline\hline
$\alpha_s$ & $\mu$ (fm$^{-1}$)& $\sigma$ ($m$b) &$\eta/s$ (Eq.~\eqref{eq:2})\\\hline
 0.33 & 3.226 & 1.5 & 0.387 \\
 0.33 & 2.265 & 3.0 & 0.232 \\
 0.33 & 1.602 & 6.0 & 0.156 \\
 0.48 & 1.800 & 10  & 0.087 \\  \hline
\end{tabular}
\end{table}

The two- and multiparticle correlations are used in the current work. The framework for the cumulants is described in Ref.~\cite{PhysRevC.83.044913}. The formalism of the multiparticle azimuthal correlations is evaluated as follows: 
\begin{equation}
\begin{split}
\label{eq:3}
 \left\langle v_n^2\right\rangle&=  \left\langle\left\langle\{2\}_{n}\right\rangle\right\rangle=\left\langle\left\langle e^{in\left(\phi_{i}-\phi_{j}\right)}\right\rangle\right\rangle,\\
\rm asc_{nm,n+m} &=\left\langle\left\langle\{3\}_{n, m}\right\rangle\right\rangle=\left\langle\left\langle e^{in\phi_i+im\phi_j-i(n+m) \phi_k}\right\rangle \right\rangle,\\
\left\langle v_n^2 v_m^2\right\rangle &= \left\langle\left\langle\{4\}_{n, m}\right\rangle\right\rangle=\left\langle\left\langle e^{in\left(\phi_{i}-\phi_{j}\right)+im\left(\phi_{k}-\phi_{m}\right)}\right\rangle\right\rangle.
\end{split}
\end{equation}
Double angular brackets $\left\langle\left\langle \right\rangle\right\rangle$ indicate that the averaging procedure has been performed on an ensemble of events with similar centrality. The non-linear coupling coefficients and  four-particle symmetric cumulants are defined as:
\begin{equation}\begin{split}\label{eq:4}
\chi_{n+m}&=\frac{asc_{nm,n+m}}{\left\langle v_n^2 v_m^2\right\rangle},\\
\operatorname{sc}_{n, m}\{4\}
&= \left\langle v_n^2 v_m^2\right\rangle-\left\langle v_n^2\right\rangle\left\langle v_m^2\right\rangle. 
\end{split}\end{equation}
To scale out the dependence on the single particle flow harmonics, the normalized symmetric cumulant is defined as follows:
\begin{equation}\begin{split}\label{eq:5}
\operatorname{nsc}_{n,m}\{4\}=\frac{\operatorname{sc}_{n,m}\{4\}}{\left\langle v_n^2 \right\rangle \left\langle v_m^2 \right\rangle}.
\end{split}\end{equation}

We analyze a three-particle correlation between triangular flow $v_3$ and mean transverse momentum $[p_{T}]$, $\left\langle v_3^2 \delta p_{\mathrm{T}}\right\rangle$,
\begin{equation}\label{eq:3}\begin{split}
\rho(v_n^2,[p_T]) = \frac{\langle v_n^2 \delta \pT \rangle}{\sqrt{\left\langle(\delta v_n^2)^2\right\rangle\left\langle\delta p_{\mathrm{T}} \delta p_{\mathrm{T}}\right\rangle}}
\end{split}
\end{equation}
Where $\langle v_n^2 \delta p_T \rangle=\langle \cos[n(\phi_i-\phi_j)] (p_{T,k} - \langle [p_T]\rangle)\rangle_{i,j,k}$. where the indices $i$, $j$ and $k$ loop over distinct particles to account for all unique triplets, and the $\langle \rangle$ denotes average over events.

In order to suppress short-range $``$non-flow" correlations arising from jet fragmentation and resonance decays, pseudorapidity gaps are often explicitly required between the particles in each pair in well-developed sub-event cumulant methods~\cite{Jia:2017hbm}. The standard cumulant methods are contaminated by non-flow correlations over the the low multiplicity region in A+A collisions. In the two-subevent cumulant method, the tracks are divided into two subevents according to $-2 < \eta_a < 0$ and $0 \leq \eta_{b}< 2$. Only hardons with 0.2 $< p_{\rm T} <$ 2 GeV/$c$ are chosen. The simulated events are binned into classes defined by the number of participant nucleons $N_{\mathrm{part}}$.

\textit{Results and discussions.---}\label{sec:2}
We focus on multiparticle cumulant calculations, which exhibit minimal sensitivity to hydrodynamic response variations. These observables are systematically calculated for different nuclear parameter configurations presented in Tab.~\ref{tab:1}. $^{96}$Ru and $^{96}$Zr in the full parameterizations of nuclear parameters are labeled Case1 and Case5 where their statistics are shown in black and red uncertainty bands, respectively, in all subsequent figures. 

\begin{figure}[!h]
\includegraphics[width=1\linewidth]{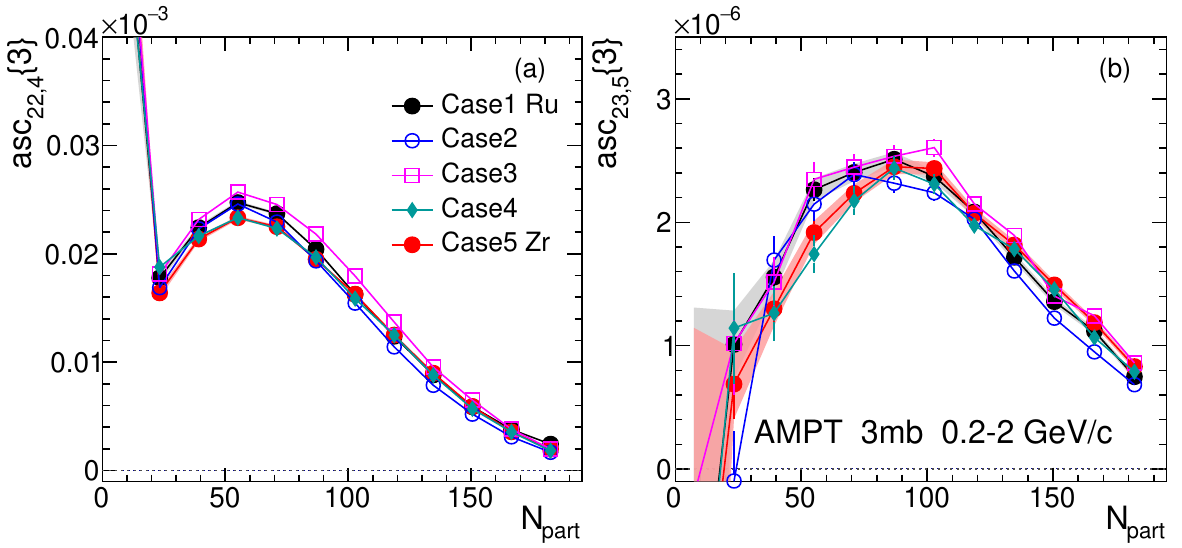}
\vspace*{-.3cm}
\caption{Asymmetric cumulant ${\rm asc}_{nm,n+m}\{3\}$ for $n,m=2,2$ and $2,3$ representing the influence of nuclear structures as a function of $N_{\rm{part}}$ in 0.2 $< p_{\rm T} <$ 2 GeV/$c$ in isobar collisions. The effects of various Woods-Saxon parameters in Tab.~\ref{tab:1} are shown.}
\label{fig:1}
\end{figure}
\begin{figure}[!h]
\includegraphics[width=1\linewidth]{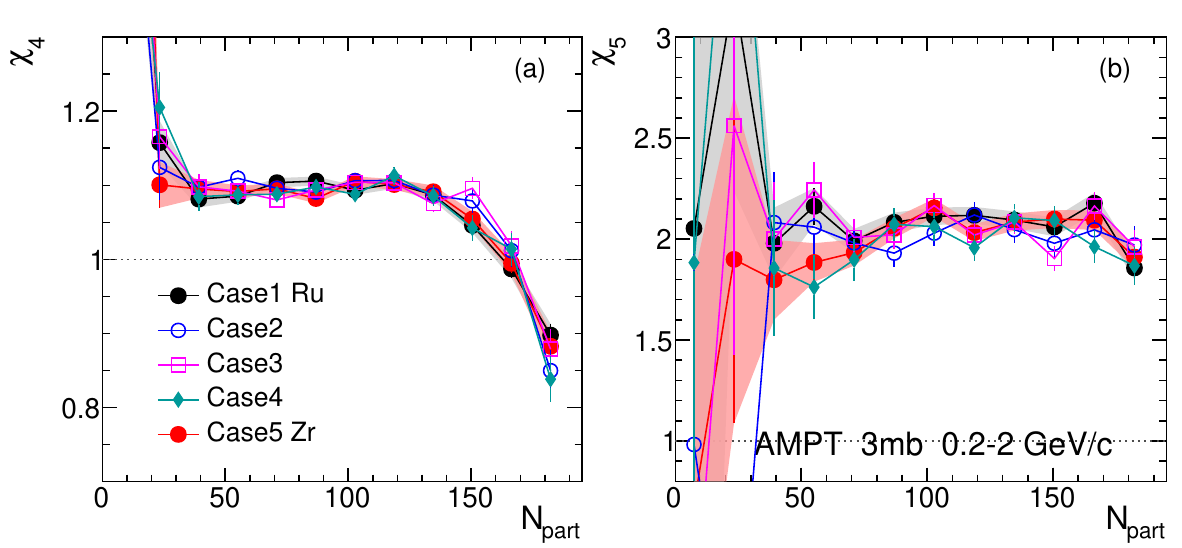}
\vspace*{-.3cm}
\caption{Nonlinear coupling coefficients $\chi_4$ (left panel) and $\chi_5$ (right panel) representing the influence of nuclear structures as a function of $N_{\rm{part}}$ in 0.2 $< p_{\rm T} <$ 2 GeV/c in isobar collisions. The effects of various Woods-Saxon parameters in Tab.~\ref{tab:1} are shown.}
\label{fig:2}
\end{figure}

The three-particle asymmetric cumulants are sensitive to correlations involving both the flow magnitude and flow phase~\cite{Jia:2017hbm}. Figure~\ref{fig:1} presents the predicted values of ${\rm asc}_{nm,n+m}\{3\}$ for the cases $n,m=2,2$ and $2,3$, which correspond to the impact of nuclear structure as a function of $N_{\rm{part}}$ in 0.2 $< p_{\rm T} <$ 2 GeV/c. The results exhibit a visiable sensitivity to nuclear structure effects, as shown by their characteristic dependence on collision centrality. Notably, the values of $\rm asc_{nm,n+m}\{3\}$ remain positive throughout the entire centrality interval considered. The correlation is weak in the central collisions, increases rapidly with a pronounced peak in the mid-central collisions, and then decreases towards more peripheral collisions. 

The ${\rm asc}_{nm,n+m}\{3\}$ is directly related to the non-linear mode mixing effects between different-order flow vector, characterized by $V_4=V_{4 \mathrm{L}}+\chi_4\left(V_2\right)^2$ and $V_5=V_{5 \mathrm{L}}+\chi_5 V_2 V_3$~\cite{Jia:2022qrq,Zhao:2022uhl,Lu:2023fqd,Magdy:2022ize,Pei:2024wsy}. 
The nonlinear coupling coefficients $\chi_4$ and $\chi_5$ determine the coupling strength between the elliptic and quadrangular flow generated during the QGP expansion, as shown in Fig.~\ref{fig:2}. Thus, they do indeed probe the transport and hadronization properties of the QGP. They show a weak centrality dependence, with only a decrease in the 0-5\% centrality interval, consistent with recent measurements~\cite{STAR:2020gcl,ALICE:2017fcd} in $^{197}$Au+$^{197}$Au collisions at RHIC and $^{208}$Pb+$^{208}$Pb collisions at the LHC. The possible existence of hexadecapole deformation $\beta_4$ in isobar collisions could also be tested using nonlinear coupling coefficients in the future. The detailed effects of nuclear structure parameters through isobar ratios, $R_{\mathcal{O}}\left(\mathrm{N_{part}}\right)=\mathcal{O}_{\mathrm{Ru}}\left(\mathrm{N_{part}}\right)/\mathcal{O}_{\mathrm{Zr}}\left(\mathrm{N_{part}}\right)$, can be accessed extensively~\cite{Jia:2022qrq}.

\begin{figure}[!b]
\includegraphics[width=1\linewidth]{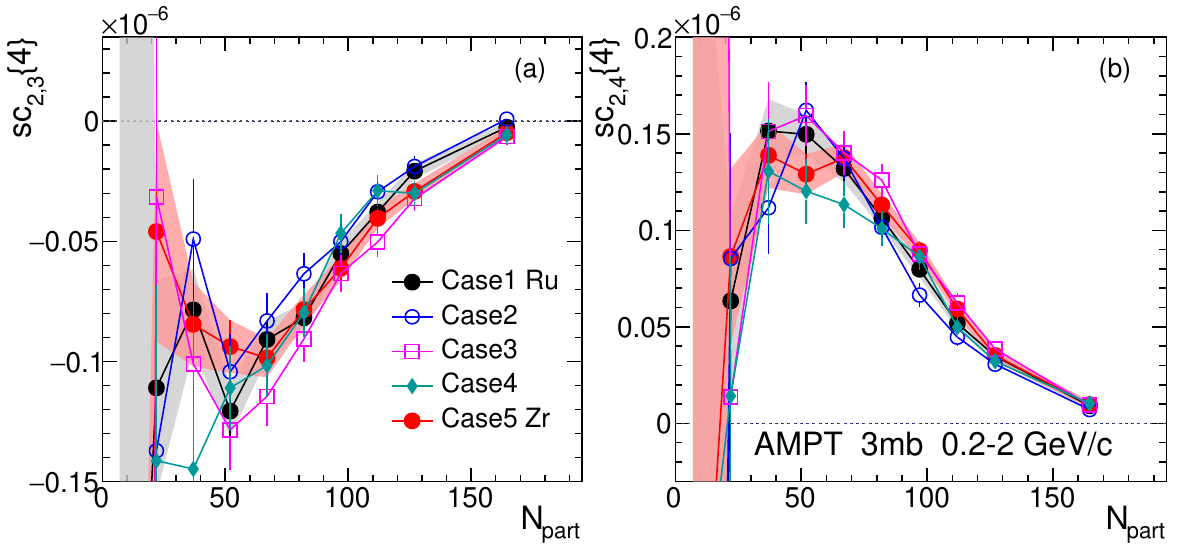}
\includegraphics[width=1\linewidth]{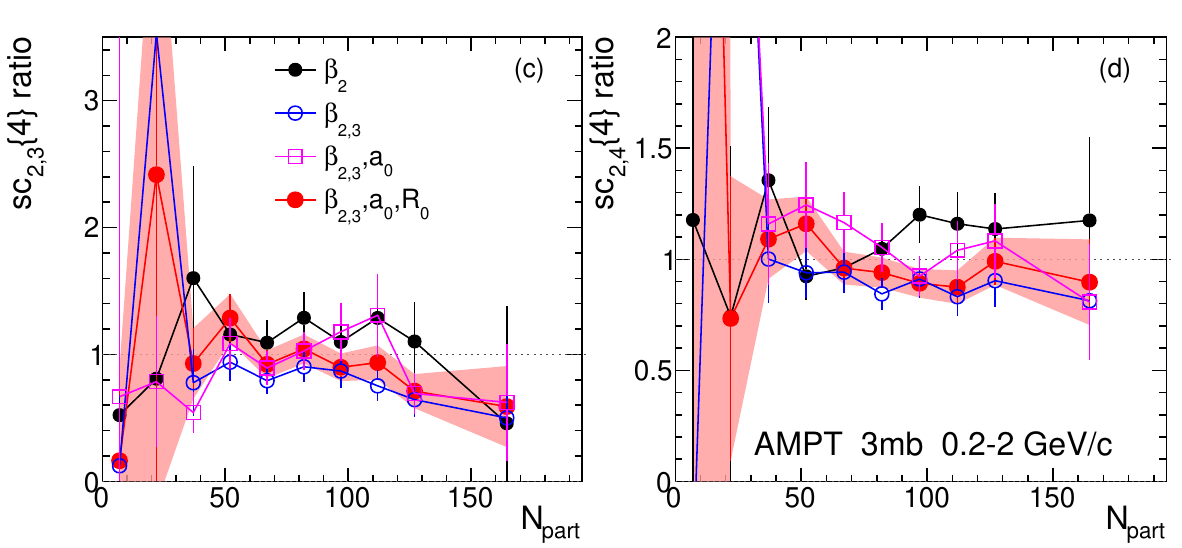}
\vspace*{-.3cm}
\caption{Symmetric cumulant ${\rm sc}_{n,m}\{4\}$ (a, b) for $n,m=2,3$ and $2,4$ and the ratios between $^{96}$Ru+$^{96}$Ru and $^{96}$Zr+$^{96}$Zr (c, d) representing the influence of nuclear structures as a function of $N_{\rm{part}}$ in 0.2 $< p_{\rm T} <$ 2 GeV/$c$. The effects of various Woods-Saxon parameters in Tab.~\ref{tab:1} are shown.}
\label{fig:3}
\end{figure}

Four-particle symmetric cumulants ${\rm sc}_{n,m}\{4\}$ quantify the lowest-order correlation between $v_n$ and $v_m$. Figure~\ref{fig:3} illustrates the symmetric cumulant ${\rm sc}_{n,m}\{4\}$ for the harmonic orders $n,m=2,3$ and $2,4$ (top panels), along with their corresponding isobar ratios (bottom panels) plotted as functions of $N_{\rm part}$. These results reveal distinct correlation patterns. Specifically, a consistently negative correlation between the flow coefficients $v_2$ and $v_3$, represented by ${\rm sc}_{2,3}\{4\}$, and a positive correlation between $v_2$ and $v_4$, denoted by ${\rm sc}_{2,4}\{4\}$ are observed across all collision modes studied and throughout the entire range of $N_{\rm part}$. Since hydrodynamic response effects are expected to be effectively canceled in the ratios between the two isobars, the observed differences can be attributed directly to variations in nuclear structure. As depicted in Fig.~\ref{fig:3} (c) and (d), these ratios explicitly reflect the influence of the nuclear structures in $^{96}$Ru and $^{96}$Zr nuclei. In particular, a difference in octupole deformation parameter $\beta_3$ leads to a suppression of approximately 50\% in the value of ${\rm sc}_{2,3}\{4\}$ in central collisions, while effects arising from neutron skin thickness and nuclear radius are effectively negated. Furthermore, the ratios of ${\rm sc}_{2,4}\{4\}$ exhibit a more complex dependence on nuclear structure parameters: the quadrupole deformation $\beta_2$ enhances the ${\rm sc}_{2,4}\{4\}$ in central collisions, whereas the octupole deformation $\beta_3$ induces an opposing effect. Additionally, an increase in neutron skin thickness elevates  ${\rm sc}_{2,4}\{4\}$ in peripheral collisions, with the nuclear radius exerting a compensatory influence. 

\begin{figure}[!h]
\includegraphics[width=1\linewidth]{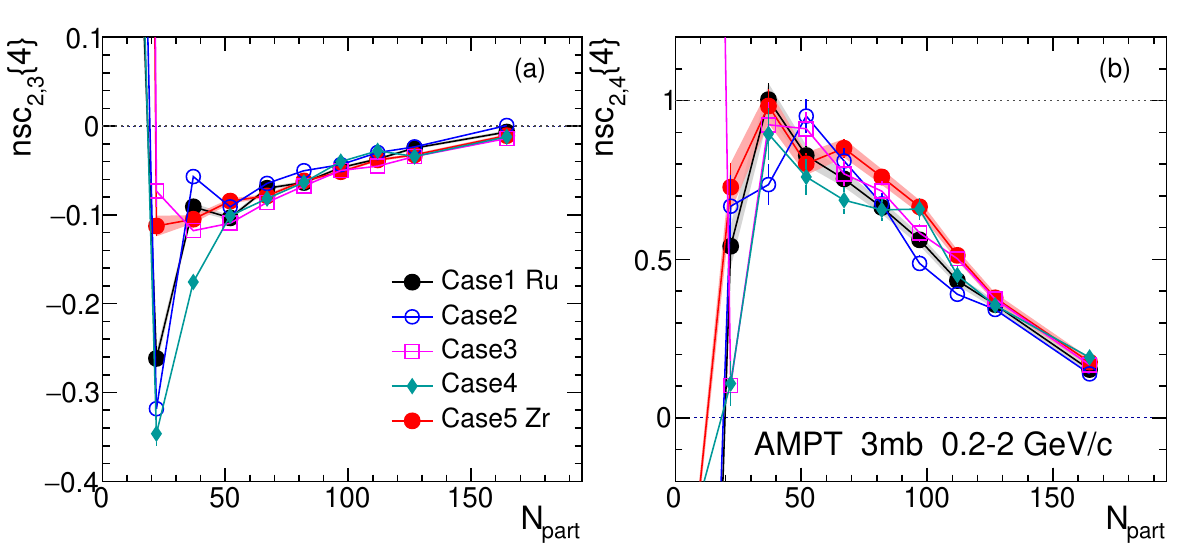}
\includegraphics[width=1\linewidth]{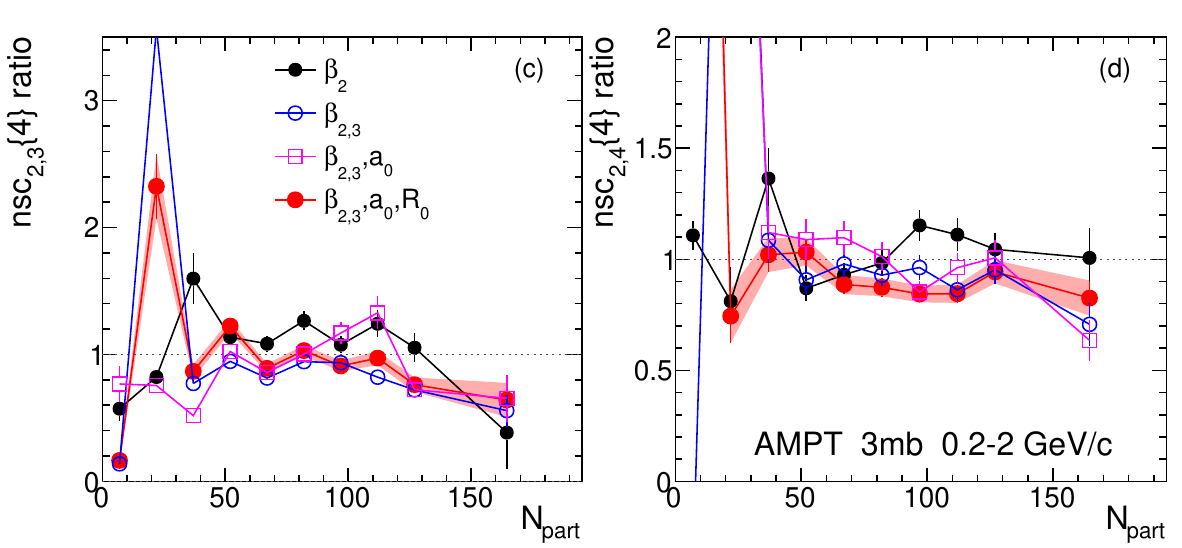}
\vspace*{-.3cm}
\caption{Normalized symmetric cumulant ${\rm nsc}_{n,m}\{4\}$ (a,b) for $n,m=2,3$ and $2,4$ and the ratios between Ru+Ru and Zr+Zr (c,d) representing the influence of nuclear structures as a function of $N_{\rm{part}}$ in 0.2 $< p_{\rm T} <$ 2 GeV/c. The effects of various Woods-Saxon parameters in Tab.~\ref{tab:1} are shown.}
\label{fig:4}
\end{figure}

To further elucidate the strength of the correlation and to disentangle the influence arising from the magnitudes of the flow harmonics, we examine the normalized cumulants ${\rm nsc}_{2,3}\{4\}$ and ${\rm nsc}_{2,4}\{4\}$, as depicted in Fig.~\ref{fig:4}. Both ${\rm nsc}_{2,3}\{4\}$ and ${\rm nsc}_{2,4}\{4\}$ exhibit a centrality dependence analogous to that observed in the unnormalized cumulants. Specifically, ${\rm nsc}_{2,3}\{4\}$ remains negative, whereas ${\rm nsc}_{2,4}\{4\}$ retains a positive value across the entire centrality range. The corresponding isobar ratios reveal distinct modification patterns: the ${\rm nsc}_{2,3}\{4\}$ ratios demonstrate enhancement and suppression effects governed by the deformation parameters $\beta_2$ and $\beta_3$ in central collisions, respectively. Additionally, a compensatory interaction between the neutron skin thickness and nuclear radius is identified, consistent with prior observations. Overall, these trends indicate suppression in central collisions. In contrast, ${\rm nsc}_{2,4}\{4\}$ manifests  more pronounced sensitivity to nuclear structure parameters. The quadrupole deformation $\beta_2$ amplifies the ratios in central collisions, whereas the octupole deformation $\beta_3$ diminishes them in peripheral collisions. A clear reduction behavior of the neutron skin thickness contribution is also evident. Incorporating the effects of all four WS parameters, we briefly summarize the nuclear structure contributions inferred from the AMPT model results as follows: 1) Nuclear deformation effects predominantly manifest in central collisions, with $\beta_{2,\mathrm{Ru}}$ and $\beta_{3,\mathrm{Zr}}$ entering opposing influences; 2) the diffuseness parameter $a_0$ and the half-density radius $R_0$ contribute substantially to the reduction of the cumulant ratios. In general, the predicted centrality dependence of the ${\rm sc}_{n,m}\{4\}$ ratios presented in Fig.~\ref{fig:3} aligns with
preliminary findings from the STAR experiment~\cite{chunjian} and the nomarlized cumulants ${\rm nsc}_{n,m}\{4\}$ shown in Fig.~\ref{fig:4} are amenable to precise experimental investigation.
\begin{figure}[!h]
\includegraphics[width=1\linewidth]{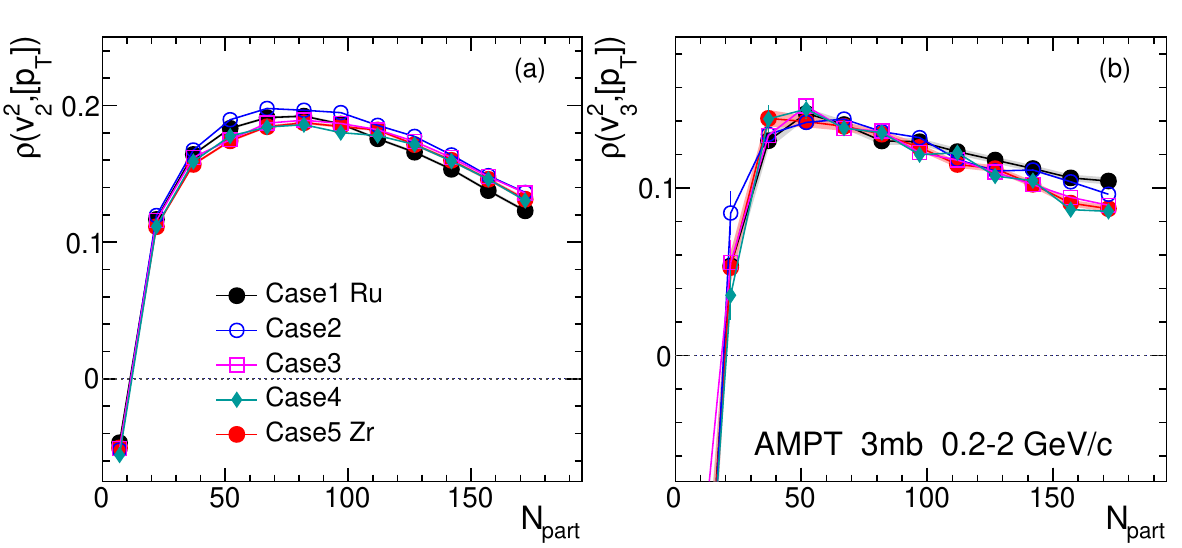}
\vspace*{-.3cm}
\includegraphics[width=1\linewidth]{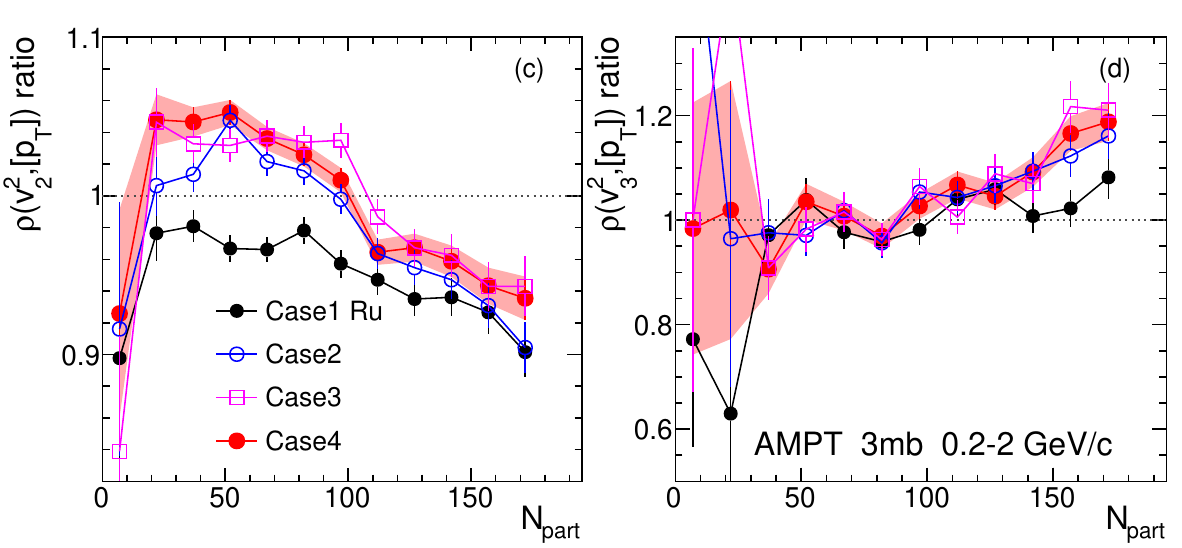}
\caption{Pearson correlation coefficient $\rho(v_n^2,[p_{\rm T}])$ (a,b) for $n=2,3$ and the ratios between Ru+Ru and Zr+Zr (c,d) representing the influence of nuclear structures as a function of $N_{\rm{part}}$ in 0.2 $< p_{\rm T} <$ 2 GeV/$c$. The effects of various Woods-Saxon parameters in Tab.~\ref{tab:1} are shown.}
\label{fig:5}
\end{figure}

The calculation of the pearson correlation coefficient $\rho_n$ reveals the significant influences of nuclear structure inputs on the shape and size correlations quantified by a three-particle correlator. 
Prior investigations have examined the effect of quadrupole deformation $\beta_2$ on the correlations between $v_2$ and the mean transverse momentum $[p_{\rm T}]$ in $^{238}$U+$^{238}$U collisions~\cite{Jia:2021wbq}. Recent experimental results from the STAR Collaboration have further indicated the presence of modest octupole deformations in $^{238}$U~\cite{STAR:2025vbp}. Nevertheless, the combined influence of nuclear deformation effects, including both quadrupole and octupole components on $\rho_n$ has not yet been systematically explored in isobar collisions using transport model frameworks. 

In this study, we conducted a dedicated analysis involving a stepwise comparison to isolate the contributions of various nuclear structure parameters to $\rho_n$ as illustrated in Fig.~\ref{fig:5}. We observed that $\rho_2$ decreases steadily, which reflects the large prolate deformation $\beta_2$ characteristic of $^{96}$Ru. This deformation exerts the most substantial influence in central collisions, although its effects persist across other centrality classes. Conversely, $\rho_3$ increases continuously, only reflecting the large octupole deformation $\beta_3$ of $^{96}$Zr. Notably, these trends are consistent with observations in the heavy nucleus $^{238}$U+$^{238}$U collision system reported by the STAR Collaboration~\cite{STAR:2024wgy,STAR:2025vbp,Zhang:2025hvi}. Furthermore, our analysis indicates that the effects arising from neutron skin thickness and nuclear size differences are negligible, a conclusion that is anticipated to extend to hydrodynamic modeling approaches as well. The transport model results present here thus provide an essential baseline comparison with STAR experimental measurements of $v_n-[p_{\rm T}]$ correlations. In particular, they facilitate the disentanglement of competing influences from quadrupole and octupole deformations, which have hitherto remained unexplored in isobar collision systems.

\begin{figure}[t]
\includegraphics[width=1\linewidth]{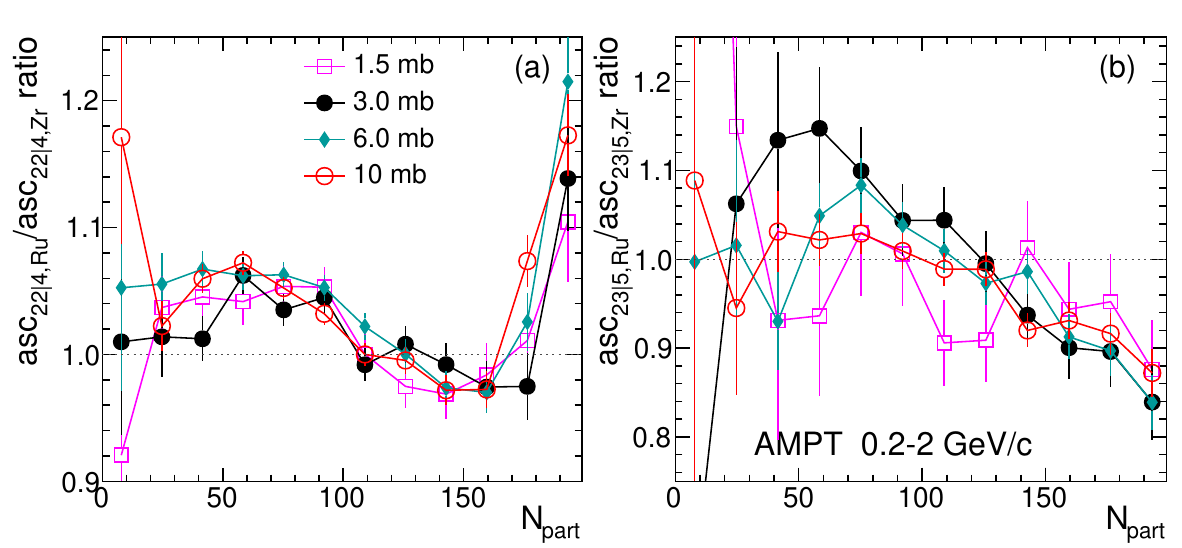}
\vspace*{-.3cm}
\caption{Ratios of the three-particle correlation asymmetric cumulant ${\rm asc}_{nm,n+m}\{3\}$ (a,b) for $n,m=2,2$ and $2,3$ between $^{96}$Ru+$^{96}$Ru and $^{96}$Zr+$^{96}$Zr, in particular, different values of the partonic cross sections as a function of $N_{\rm{part}}$ in 0.2 $< p_{\rm T} <$ 2 GeV/$c$.}
\label{fig:6}
\end{figure}

\begin{table}[t]
\caption{Ratios of the three-particle correlation asymmetric cumulant ${\rm asc}_{nm,n+m}\{3\}$ between $^{96}$Ru+$^{96}$Ru and $^{96}$Zr+$^{96}$Zr for different observables in 0--10\% centrality with four partonic cross listed in Tab.~\ref{tab:2}.}
\label{TableIII}
\centering
\scriptsize
\begingroup
\renewcommand{\arraystretch}{1.2}
\begin{tabular}{lcccc}
\hline\hline
Obs. & 1.5 mb & 3.0 mb & 6.0 mb & 10.0 mb \\\hline
asc$_{22|4}$ & 1.033 $\pm$ 0.035 & 1.029 $\pm$ 0.029 & 1.070 $\pm$ 0.025 & 1.073 $\pm$ 0.023 \\
asc$_{23|5}$ & 0.924 $\pm$ 0.054 & 0.879 $\pm$ 0.039 & 0.883 $\pm$ 0.028 & 0.906 $\pm$ 0.024 \\\hline
\end{tabular}
\endgroup
\end{table}

It is important to note that the final state effect are expected to canceled when considering the ratios between isobars. Extending our previous work~\cite{Zhang:2022fou}, we systematically carried out the checks for the shear viscosity effect to correlated fluctuations of flow harmonics. The effect of the partonic cross section on the ratios of the asymmetric cumulant ${\rm asc}_{nm,n+m}\{3\}$ between $^{96}$Ru+$^{96}$Ru and $^{96}$Zr+$^{96}$Zr collisions are shown in Fig.~\ref{fig:6}. The deviation of these ratios from unity shown in nonmonotonic behavior is related to the difference in the collective nuclear structure between $^{96}$Ru and $^{96}$Zr nuclei in terms of quadrupole deformation $\beta_2$, octupole deformation $\beta_3$, neutron skin $a_0$, and nuclear radius $R_0$, which has been extensively discussed in detail in~\cite{Jia:2022qrq}. The values of the ratios in the 0-10\% centrality, as shown in Tab.~\ref{TableIII}, remains almost unchanged, implying that the isobar ratios of the multiparticle correlations are insensitive to the medium properties in the final state. We emphasize that this analysis can be performed directly if our prediction is confirmed by the experiment and other hydrodynamical model calculations.

In addition, we have cross-checked the shear viscosity effect on the ratio of non-linear coupling coefficients $\chi_n$, symmetric cumulant $\rm sc_{n,m}\{4\}$ and the normalized symmetric cumulant $\rm nsc_{n,m}\{4\}$ for $n,m=2,3$ and $2,4$. We indeed see that these results imply a negligible shear viscosity effect even with larger errors within current model statistics. Nevertheless, the experimental datasets collected by the STAR Collaboration comprise approximately two billion events for each isobar species~\cite{STAR:2021mii}, offering a distinctive opportunity for the aforementioned measurements that require substantial statistical power.

\textit{Summary.---}
In summary, we have conducted a systematic investigation into the impact of nuclear structure parameters on long-range multiparticle correlations and their ratios in $^{96}$Ru+$^{96}$Ru and $^{96}$Zr+$^{96}$Zr isobar collisions at $\sqrtsnn$ = 200 GeV employing the AMPT model. Our analysis predicts distinct behaviors for asymmetric cumulant, non-linear mode coefficients, (normalized-)symmetric cumulants, and the correlated fluctuations between $v_n$ and $[p_{\rm T}]$. Notably, we observe significant deviations of these ratios from unity, which are attributed to the difference between the collective nuclear structures of the $^{96}$Ru and $^{96}$Zr nuclei. The quadrupole deformation parameter $\beta_2$ and the octupole deformation parameter $\beta_3$ exhibit a pronounced influence, whereas the effects of nuclear radius $R_0$ and neutron skin thickness ($a_0$) tend to largely offset one another. Moreover, our results indicate that shear viscosity has a negligible effect on the correlations among different flow harmonics. We anticipate that these comprehensive findings will serve as essential theoretical inputs for the STAR experimental measurements~\cite{chunjian,ChunjianQM} and provide a robust framework for quantifying the characteristics of initial-state fluctuations. These multiple observables could also help us to constrain the nuclear parameters in a simultaneous manner. Nevertheless, naturally applying such a possibility to other isobars in the nuclear chart or isobar-like collisions such as $^{16}$O+$^{16}$O and $^{20}$Ne+$^{20}$Ne, could further advance the understanding of many-body properties in nuclear clusters particularly in light of emerging modern \textit{ab initio} theoretical methodologies~\cite{Giacalone:2024luz,Giacalone:2024ixe,Zhang:2024vkh,Zhao:2024feh,YuanyuanWang:2024sgp,Wang:2024ulq,Liu:2023gun,Lu:2025cni}.

\textit {Acknowledgements.---}
The authors thank Somadutta Bhatta and Lumeng Liu for valuable comments. This work is founded by the National Key Research and Development Program of China under Contract No. 2024YFA1612600 and No. 2022YFA1604900, the National Natural Science Foundation of China (NSFC) under Contract No. 12025501 and No. 12547102, the U.S. Department of Energy, Office of Science, Office of Nuclear Physics, under DOE Awards No. DE-SC0024602.
\bibliography{ref}

\begin{thebibliography}{99}%
\makeatletter
\providecommand \@ifxundefined [1]{%
 \@ifx{#1\undefined}
}%
\providecommand \@ifnum [1]{%
 \ifnum #1\expandafter \@firstoftwo
 \else \expandafter \@secondoftwo
 \fi
}%
\providecommand \@ifx [1]{%
 \ifx #1\expandafter \@firstoftwo
 \else \expandafter \@secondoftwo
 \fi
}%
\providecommand \natexlab [1]{#1}%
\providecommand \enquote  [1]{``#1''}%
\providecommand \bibnamefont  [1]{#1}%
\providecommand \bibfnamefont [1]{#1}%
\providecommand \citenamefont [1]{#1}%
\providecommand \href@noop [0]{\@secondoftwo}%
\providecommand \href [0]{\begingroup \@sanitize@url \@href}%
\providecommand \@href[1]{\@@startlink{#1}\@@href}%
\providecommand \@@href[1]{\endgroup#1\@@endlink}%
\providecommand \@sanitize@url [0]{\catcode `\\12\catcode `\$12\catcode `\&12\catcode `\#12\catcode `\^12\catcode `\_12\catcode `\%12\relax}%
\providecommand \@@startlink[1]{}%
\providecommand \@@endlink[0]{}%
\providecommand \url  [0]{\begingroup\@sanitize@url \@url }%
\providecommand \@url [1]{\endgroup\@href {#1}{\urlprefix }}%
\providecommand \urlprefix  [0]{URL }%
\providecommand \Eprint [0]{\href }%
\providecommand \doibase [0]{http://dx.doi.org/}%
\providecommand \selectlanguage [0]{\@gobble}%
\providecommand \bibinfo  [0]{\@secondoftwo}%
\providecommand \bibfield  [0]{\@secondoftwo}%
\providecommand \translation [1]{[#1]}%
\providecommand \BibitemOpen [0]{}%
\providecommand \bibitemStop [0]{}%
\providecommand \bibitemNoStop [0]{.\EOS\space}%
\providecommand \EOS [0]{\spacefactor3000\relax}%
\providecommand \BibitemShut  [1]{\csname bibitem#1\endcsname}%
\let\auto@bib@innerbib\@empty
\bibitem [{\citenamefont {Ollitrault}(1992)}]{Ollitrault:1992bk}%
  \BibitemOpen
  \bibfield  {author} {\bibinfo {author} {\bibfnamefont {J.-Y.}\ \bibnamefont {Ollitrault}},\ }\href {\doibase 10.1103/PhysRevD.46.229} {\bibfield  {journal} {\bibinfo  {journal} {Phys. Rev. D}\ }\textbf {\bibinfo {volume} {46}},\ \bibinfo {pages} {229} (\bibinfo {year} {1992})}\BibitemShut {NoStop}%
\bibitem [{\citenamefont {Teaney}\ \emph {et~al.}(2001)\citenamefont {Teaney}, \citenamefont {Lauret},\ and\ \citenamefont {Shuryak}}]{Teaney:2000cw}%
  \BibitemOpen
  \bibfield  {author} {\bibinfo {author} {\bibfnamefont {D.}~\bibnamefont {Teaney}}, \bibinfo {author} {\bibfnamefont {J.}~\bibnamefont {Lauret}}, \ and\ \bibinfo {author} {\bibfnamefont {E.~V.}\ \bibnamefont {Shuryak}},\ }\href {\doibase 10.1103/PhysRevLett.86.4783} {\bibfield  {journal} {\bibinfo  {journal} {Phys. Rev. Lett.}\ }\textbf {\bibinfo {volume} {86}},\ \bibinfo {pages} {4783} (\bibinfo {year} {2001})},\ \Eprint {http://arxiv.org/abs/nucl-th/0011058} {arXiv:nucl-th/0011058} \BibitemShut {NoStop}%
\bibitem [{\citenamefont {Voloshin}\ \emph {et~al.}(2010)\citenamefont {Voloshin}, \citenamefont {Poskanzer},\ and\ \citenamefont {Snellings}}]{Voloshin:2008dg}%
  \BibitemOpen
  \bibfield  {author} {\bibinfo {author} {\bibfnamefont {S.~A.}\ \bibnamefont {Voloshin}}, \bibinfo {author} {\bibfnamefont {A.~M.}\ \bibnamefont {Poskanzer}}, \ and\ \bibinfo {author} {\bibfnamefont {R.}~\bibnamefont {Snellings}},\ }\href {\doibase 10.1007/978-3-642-01539-7_10} {\bibfield  {journal} {\bibinfo  {journal} {Landolt-Bornstein}\ }\textbf {\bibinfo {volume} {23}},\ \bibinfo {pages} {293} (\bibinfo {year} {2010})},\ \Eprint {http://arxiv.org/abs/0809.2949} {arXiv:0809.2949 [nucl-ex]} \BibitemShut {NoStop}%
\bibitem [{\citenamefont {Gale}\ \emph {et~al.}(2013)\citenamefont {Gale}, \citenamefont {Jeon},\ and\ \citenamefont {Schenke}}]{Gale:2013da}%
  \BibitemOpen
  \bibfield  {author} {\bibinfo {author} {\bibfnamefont {C.}~\bibnamefont {Gale}}, \bibinfo {author} {\bibfnamefont {S.}~\bibnamefont {Jeon}}, \ and\ \bibinfo {author} {\bibfnamefont {B.}~\bibnamefont {Schenke}},\ }\href {\doibase 10.1142/S0217751X13400113} {\bibfield  {journal} {\bibinfo  {journal} {Int. J. Mod. Phys.}\ }\textbf {\bibinfo {volume} {A28}},\ \bibinfo {pages} {1340011} (\bibinfo {year} {2013})},\ \Eprint {http://arxiv.org/abs/1301.5893} {arXiv:1301.5893 [nucl-th]} \BibitemShut {NoStop}%
\bibitem [{\citenamefont {Heinz}\ and\ \citenamefont {Snellings}(2013)}]{Heinz:2013th}%
  \BibitemOpen
  \bibfield  {author} {\bibinfo {author} {\bibfnamefont {U.}~\bibnamefont {Heinz}}\ and\ \bibinfo {author} {\bibfnamefont {R.}~\bibnamefont {Snellings}},\ }\href {\doibase 10.1146/annurev-nucl-102212-170540} {\bibfield  {journal} {\bibinfo  {journal} {Ann. Rev. Nucl. Part. Sci.}\ }\textbf {\bibinfo {volume} {63}},\ \bibinfo {pages} {123} (\bibinfo {year} {2013})},\ \Eprint {http://arxiv.org/abs/1301.2826} {arXiv:1301.2826 [nucl-th]} \BibitemShut {NoStop}%
\bibitem [{\citenamefont {Romatschke}\ and\ \citenamefont {Romatschke}(2019)}]{Romatschke:2017ejr}%
  \BibitemOpen
  \bibfield  {author} {\bibinfo {author} {\bibfnamefont {P.}~\bibnamefont {Romatschke}}\ and\ \bibinfo {author} {\bibfnamefont {U.}~\bibnamefont {Romatschke}},\ }\href {\doibase 10.1017/9781108651998} {\emph {\bibinfo {title} {{Relativistic Fluid Dynamics In and Out of Equilibrium}}}},\ Cambridge Monographs on Mathematical Physics\ (\bibinfo  {publisher} {Cambridge University Press},\ \bibinfo {year} {2019})\ \Eprint {http://arxiv.org/abs/1712.05815} {arXiv:1712.05815 [nucl-th]} \BibitemShut {NoStop}%
\bibitem [{\citenamefont {Alqahtani}\ and\ \citenamefont {Ollitrault}(2026)}]{Alqahtani:2025wan}%
  \BibitemOpen
  \bibfield  {author} {\bibinfo {author} {\bibfnamefont {M.}~\bibnamefont {Alqahtani}}\ and\ \bibinfo {author} {\bibfnamefont {J.-Y.}\ \bibnamefont {Ollitrault}},\ }\href {\doibase 10.1016/j.physletb.2025.140066} {\bibfield  {journal} {\bibinfo  {journal} {Phys. Lett. B}\ }\textbf {\bibinfo {volume} {872}},\ \bibinfo {pages} {140066} (\bibinfo {year} {2026})},\ \Eprint {http://arxiv.org/abs/2507.20315} {arXiv:2507.20315 [nucl-th]} \BibitemShut {NoStop}%
\bibitem [{\citenamefont {Teaney}\ and\ \citenamefont {Yan}(2012)}]{Teaney:2012ke}%
  \BibitemOpen
  \bibfield  {author} {\bibinfo {author} {\bibfnamefont {D.}~\bibnamefont {Teaney}}\ and\ \bibinfo {author} {\bibfnamefont {L.}~\bibnamefont {Yan}},\ }\href {\doibase 10.1103/PhysRevC.86.044908} {\bibfield  {journal} {\bibinfo  {journal} {Phys. Rev. C}\ }\textbf {\bibinfo {volume} {86}},\ \bibinfo {pages} {044908} (\bibinfo {year} {2012})},\ \Eprint {http://arxiv.org/abs/1206.1905} {arXiv:1206.1905 [nucl-th]} \BibitemShut {NoStop}%
\bibitem [{\citenamefont {Deng}\ \emph {et~al.}(2024)\citenamefont {Deng}, \citenamefont {Fang},\ and\ \citenamefont {Ma}}]{Deng:2024}%
  \BibitemOpen
  \bibfield  {author} {\bibinfo {author} {\bibfnamefont {X.-G.}\ \bibnamefont {Deng}}, \bibinfo {author} {\bibfnamefont {D.-Q.}\ \bibnamefont {Fang}}, \ and\ \bibinfo {author} {\bibfnamefont {Y.-G.}\ \bibnamefont {Ma}},\ }\href {\doibase 10.1016/j.ppnp.2023.104095} {\bibfield  {journal} {\bibinfo  {journal} {Prog. Part. Nucl. Phys.}\ ,\ \bibinfo {pages} {104095}} (\bibinfo {year} {2024})}\BibitemShut {NoStop}%
\bibitem [{\citenamefont {Pratt}\ \emph {et~al.}(2015)\citenamefont {Pratt}, \citenamefont {Sangaline}, \citenamefont {Sorensen},\ and\ \citenamefont {Wang}}]{PhysRevLett.114.202301}%
  \BibitemOpen
  \bibfield  {author} {\bibinfo {author} {\bibfnamefont {S.}~\bibnamefont {Pratt}}, \bibinfo {author} {\bibfnamefont {E.}~\bibnamefont {Sangaline}}, \bibinfo {author} {\bibfnamefont {P.}~\bibnamefont {Sorensen}}, \ and\ \bibinfo {author} {\bibfnamefont {H.}~\bibnamefont {Wang}},\ }\href {\doibase 10.1103/PhysRevLett.114.202301} {\bibfield  {journal} {\bibinfo  {journal} {Phys. Rev. Lett.}\ }\textbf {\bibinfo {volume} {114}},\ \bibinfo {pages} {202301} (\bibinfo {year} {2015})}\BibitemShut {NoStop}%
\bibitem [{\citenamefont {Adamczyk}\ \emph {et~al.}(2015)\citenamefont {Adamczyk} \emph {et~al.}}]{STAR:2015mki}%
  \BibitemOpen
  \bibfield  {author} {\bibinfo {author} {\bibfnamefont {L.}~\bibnamefont {Adamczyk}} \emph {et~al.} (\bibinfo {collaboration} {STAR}),\ }\href {\doibase 10.1103/PhysRevLett.115.222301} {\bibfield  {journal} {\bibinfo  {journal} {Phys. Rev. Lett.}\ }\textbf {\bibinfo {volume} {115}},\ \bibinfo {pages} {222301} (\bibinfo {year} {2015})},\ \Eprint {http://arxiv.org/abs/1505.07812} {arXiv:1505.07812 [nucl-ex]} \BibitemShut {NoStop}%
\bibitem [{\citenamefont {Adam}\ \emph {et~al.}(2020)\citenamefont {Adam} \emph {et~al.}}]{STAR:2020gcl}%
  \BibitemOpen
  \bibfield  {author} {\bibinfo {author} {\bibfnamefont {J.}~\bibnamefont {Adam}} \emph {et~al.} (\bibinfo {collaboration} {STAR}),\ }\href {\doibase 10.1016/j.physletb.2020.135728} {\bibfield  {journal} {\bibinfo  {journal} {Phys. Lett. B}\ }\textbf {\bibinfo {volume} {809}},\ \bibinfo {pages} {135728} (\bibinfo {year} {2020})},\ \Eprint {http://arxiv.org/abs/2006.13537} {arXiv:2006.13537 [nucl-ex]} \BibitemShut {NoStop}%
\bibitem [{\citenamefont {Abdallah}\ \emph {et~al.}(2022{\natexlab{a}})\citenamefont {Abdallah} \emph {et~al.}}]{STAR:2022gki}%
  \BibitemOpen
  \bibfield  {author} {\bibinfo {author} {\bibfnamefont {M.}~\bibnamefont {Abdallah}} \emph {et~al.} (\bibinfo {collaboration} {STAR}),\ }\href {\doibase 10.1103/PhysRevLett.129.252301} {\bibfield  {journal} {\bibinfo  {journal} {Phys. Rev. Lett.}\ }\textbf {\bibinfo {volume} {129}},\ \bibinfo {pages} {252301} (\bibinfo {year} {2022}{\natexlab{a}})},\ \Eprint {http://arxiv.org/abs/2201.10365} {arXiv:2201.10365 [nucl-ex]} \BibitemShut {NoStop}%
\bibitem [{\citenamefont {Abdulhamid}\ \emph {et~al.}(2023)\citenamefont {Abdulhamid} \emph {et~al.}}]{STAR:2022pfn}%
  \BibitemOpen
  \bibfield  {author} {\bibinfo {author} {\bibfnamefont {M.~I.}\ \bibnamefont {Abdulhamid}} \emph {et~al.} (\bibinfo {collaboration} {STAR}),\ }\href {\doibase 10.1103/PhysRevLett.130.242301} {\bibfield  {journal} {\bibinfo  {journal} {Phys. Rev. Lett.}\ }\textbf {\bibinfo {volume} {130}},\ \bibinfo {pages} {242301} (\bibinfo {year} {2023})},\ \Eprint {http://arxiv.org/abs/2210.11352} {arXiv:2210.11352 [nucl-ex]} \BibitemShut {NoStop}%
\bibitem [{\citenamefont {Aaboud}\ \emph {et~al.}(2020)\citenamefont {Aaboud} \emph {et~al.}}]{ATLAS:2019peb}%
  \BibitemOpen
  \bibfield  {author} {\bibinfo {author} {\bibfnamefont {M.}~\bibnamefont {Aaboud}} \emph {et~al.} (\bibinfo {collaboration} {ATLAS}),\ }\href {\doibase 10.1007/JHEP01(2020)051} {\bibfield  {journal} {\bibinfo  {journal} {JHEP}\ }\textbf {\bibinfo {volume} {01}},\ \bibinfo {pages} {051} (\bibinfo {year} {2020})},\ \Eprint {http://arxiv.org/abs/1904.04808} {arXiv:1904.04808 [nucl-ex]} \BibitemShut {NoStop}%
\bibitem [{\citenamefont {Adam}\ \emph {et~al.}(2016)\citenamefont {Adam} \emph {et~al.}}]{ALICE:2016kpq}%
  \BibitemOpen
  \bibfield  {author} {\bibinfo {author} {\bibfnamefont {J.}~\bibnamefont {Adam}} \emph {et~al.} (\bibinfo {collaboration} {ALICE}),\ }\href {\doibase 10.1103/PhysRevLett.117.182301} {\bibfield  {journal} {\bibinfo  {journal} {Phys. Rev. Lett.}\ }\textbf {\bibinfo {volume} {117}},\ \bibinfo {pages} {182301} (\bibinfo {year} {2016})},\ \Eprint {http://arxiv.org/abs/1604.07663} {arXiv:1604.07663 [nucl-ex]} \BibitemShut {NoStop}%
\bibitem [{\citenamefont {Sirunyan}\ \emph {et~al.}(2019)\citenamefont {Sirunyan} \emph {et~al.}}]{CMS:2017glf}%
  \BibitemOpen
  \bibfield  {author} {\bibinfo {author} {\bibfnamefont {A.~M.}\ \bibnamefont {Sirunyan}} \emph {et~al.} (\bibinfo {collaboration} {CMS}),\ }\href {\doibase 10.1016/j.physletb.2018.11.063} {\bibfield  {journal} {\bibinfo  {journal} {Phys. Lett. B}\ }\textbf {\bibinfo {volume} {789}},\ \bibinfo {pages} {643} (\bibinfo {year} {2019})},\ \Eprint {http://arxiv.org/abs/1711.05594} {arXiv:1711.05594 [nucl-ex]} \BibitemShut {NoStop}%
\bibitem [{\citenamefont {Jia}(2025)}]{Jia:2025wey}%
  \BibitemOpen
  \bibfield  {author} {\bibinfo {author} {\bibfnamefont {J.}~\bibnamefont {Jia}},\ }\href {\doibase 10.1088/1361-6633/ae0654} {\bibfield  {journal} {\bibinfo  {journal} {Rept. Prog. Phys.}\ }\textbf {\bibinfo {volume} {88}},\ \bibinfo {pages} {092301} (\bibinfo {year} {2025})},\ \Eprint {http://arxiv.org/abs/2501.16071} {arXiv:2501.16071 [nucl-th]} \BibitemShut {NoStop}%
\bibitem [{\citenamefont {Ke}(2025)}]{Ke:2025tyv}%
  \BibitemOpen
  \bibfield  {author} {\bibinfo {author} {\bibfnamefont {W.}~\bibnamefont {Ke}},\ }\href@noop {} {\  (\bibinfo {year} {2025})},\ \Eprint {http://arxiv.org/abs/2509.09549} {arXiv:2509.09549 [nucl-th]} \BibitemShut {NoStop}%
\bibitem [{\citenamefont {Giacalone}(2023)}]{Giacalone:2023hwk}%
  \BibitemOpen
  \bibfield  {author} {\bibinfo {author} {\bibfnamefont {G.}~\bibnamefont {Giacalone}},\ }\href {\doibase 10.1140/epja/s10050-023-01200-7} {\bibfield  {journal} {\bibinfo  {journal} {Eur. Phys. J. A}\ }\textbf {\bibinfo {volume} {59}},\ \bibinfo {pages} {297} (\bibinfo {year} {2023})},\ \Eprint {http://arxiv.org/abs/2305.19843} {arXiv:2305.19843 [nucl-th]} \BibitemShut {NoStop}%
\bibitem [{\citenamefont {Zhao}\ and\ \citenamefont {Ma}(2022)}]{Zhao:2022grq}%
  \BibitemOpen
  \bibfield  {author} {\bibinfo {author} {\bibfnamefont {X.-L.}\ \bibnamefont {Zhao}}\ and\ \bibinfo {author} {\bibfnamefont {G.-L.}\ \bibnamefont {Ma}},\ }\href {\doibase 10.1103/PhysRevC.106.034909} {\bibfield  {journal} {\bibinfo  {journal} {Phys. Rev. C}\ }\textbf {\bibinfo {volume} {106}},\ \bibinfo {pages} {034909} (\bibinfo {year} {2022})},\ \Eprint {http://arxiv.org/abs/2203.15214} {arXiv:2203.15214 [nucl-th]} \BibitemShut {NoStop}%
\bibitem [{\citenamefont {Kharzeev}\ \emph {et~al.}(2022)\citenamefont {Kharzeev}, \citenamefont {Liao},\ and\ \citenamefont {Shi}}]{Kharzeev:2022hqz}%
  \BibitemOpen
  \bibfield  {author} {\bibinfo {author} {\bibfnamefont {D.~E.}\ \bibnamefont {Kharzeev}}, \bibinfo {author} {\bibfnamefont {J.}~\bibnamefont {Liao}}, \ and\ \bibinfo {author} {\bibfnamefont {S.}~\bibnamefont {Shi}},\ }\href {\doibase 10.1103/PhysRevC.106.L051903} {\bibfield  {journal} {\bibinfo  {journal} {Phys. Rev. C}\ }\textbf {\bibinfo {volume} {106}},\ \bibinfo {pages} {L051903} (\bibinfo {year} {2022})},\ \Eprint {http://arxiv.org/abs/2205.00120} {arXiv:2205.00120 [nucl-th]} \BibitemShut {NoStop}%
\bibitem [{\citenamefont {Li}\ \emph {et~al.}(2022)\citenamefont {Li}, \citenamefont {Ma}, \citenamefont {Zhang}, \citenamefont {Ma},\ and\ \citenamefont {Shou}}]{PhysRevC.106.014906}%
  \BibitemOpen
  \bibfield  {author} {\bibinfo {author} {\bibfnamefont {F.}~\bibnamefont {Li}}, \bibinfo {author} {\bibfnamefont {Y.-G.}\ \bibnamefont {Ma}}, \bibinfo {author} {\bibfnamefont {S.}~\bibnamefont {Zhang}}, \bibinfo {author} {\bibfnamefont {G.-L.}\ \bibnamefont {Ma}}, \ and\ \bibinfo {author} {\bibfnamefont {Q.}~\bibnamefont {Shou}},\ }\href {\doibase 10.1103/PhysRevC.106.014906} {\bibfield  {journal} {\bibinfo  {journal} {Phys. Rev. C}\ }\textbf {\bibinfo {volume} {106}},\ \bibinfo {pages} {014906} (\bibinfo {year} {2022})}\BibitemShut {NoStop}%
\bibitem [{\citenamefont {Shou}\ \emph {et~al.}(2015)\citenamefont {Shou}, \citenamefont {Ma}, \citenamefont {Sorensen}, \citenamefont {Tang}, \citenamefont {Videb\ae{}k},\ and\ \citenamefont {Wang}}]{Shou:2014eya}%
  \BibitemOpen
  \bibfield  {author} {\bibinfo {author} {\bibfnamefont {Q.~Y.}\ \bibnamefont {Shou}}, \bibinfo {author} {\bibfnamefont {Y.~G.}\ \bibnamefont {Ma}}, \bibinfo {author} {\bibfnamefont {P.}~\bibnamefont {Sorensen}}, \bibinfo {author} {\bibfnamefont {A.~H.}\ \bibnamefont {Tang}}, \bibinfo {author} {\bibfnamefont {F.}~\bibnamefont {Videb\ae{}k}}, \ and\ \bibinfo {author} {\bibfnamefont {H.}~\bibnamefont {Wang}},\ }\href {\doibase 10.1016/j.physletb.2015.07.078} {\bibfield  {journal} {\bibinfo  {journal} {Phys. Lett. B}\ }\textbf {\bibinfo {volume} {749}},\ \bibinfo {pages} {215} (\bibinfo {year} {2015})},\ \Eprint {http://arxiv.org/abs/1409.8375} {arXiv:1409.8375 [nucl-th]} \BibitemShut {NoStop}%
\bibitem [{\citenamefont {Giacalone}\ \emph {et~al.}(2021)\citenamefont {Giacalone}, \citenamefont {Jia},\ and\ \citenamefont {Zhang}}]{PhysRevLett.127.242301}%
  \BibitemOpen
  \bibfield  {author} {\bibinfo {author} {\bibfnamefont {G.}~\bibnamefont {Giacalone}}, \bibinfo {author} {\bibfnamefont {J.}~\bibnamefont {Jia}}, \ and\ \bibinfo {author} {\bibfnamefont {C.}~\bibnamefont {Zhang}},\ }\href {\doibase 10.1103/PhysRevLett.127.242301} {\bibfield  {journal} {\bibinfo  {journal} {Phys. Rev. Lett.}\ }\textbf {\bibinfo {volume} {127}},\ \bibinfo {pages} {242301} (\bibinfo {year} {2021})}\BibitemShut {NoStop}%
\bibitem [{\citenamefont {Nie}\ \emph {et~al.}(2023)\citenamefont {Nie}, \citenamefont {Zhang}, \citenamefont {Chen}, \citenamefont {Yi},\ and\ \citenamefont {Jia}}]{Nie:2022gbg}%
  \BibitemOpen
  \bibfield  {author} {\bibinfo {author} {\bibfnamefont {M.}~\bibnamefont {Nie}}, \bibinfo {author} {\bibfnamefont {C.}~\bibnamefont {Zhang}}, \bibinfo {author} {\bibfnamefont {Z.}~\bibnamefont {Chen}}, \bibinfo {author} {\bibfnamefont {L.}~\bibnamefont {Yi}}, \ and\ \bibinfo {author} {\bibfnamefont {J.}~\bibnamefont {Jia}},\ }\href {\doibase 10.1016/j.physletb.2023.138177} {\bibfield  {journal} {\bibinfo  {journal} {Phys. Lett. B}\ }\textbf {\bibinfo {volume} {845}},\ \bibinfo {pages} {138177} (\bibinfo {year} {2023})},\ \Eprint {http://arxiv.org/abs/2208.05416} {arXiv:2208.05416 [nucl-th]} \BibitemShut {NoStop}%
\bibitem [{\citenamefont {Jia}(2022{\natexlab{a}})}]{Jia:2021tzt}%
  \BibitemOpen
  \bibfield  {author} {\bibinfo {author} {\bibfnamefont {J.}~\bibnamefont {Jia}},\ }\href {\doibase 10.1103/PhysRevC.105.014905} {\bibfield  {journal} {\bibinfo  {journal} {Phys. Rev. C}\ }\textbf {\bibinfo {volume} {105}},\ \bibinfo {pages} {014905} (\bibinfo {year} {2022}{\natexlab{a}})},\ \Eprint {http://arxiv.org/abs/2106.08768} {arXiv:2106.08768 [nucl-th]} \BibitemShut {NoStop}%
\bibitem [{\citenamefont {Xu}\ \emph {et~al.}(2023)\citenamefont {Xu}, \citenamefont {Zhao}, \citenamefont {Li}, \citenamefont {Zhou}, \citenamefont {Chen},\ and\ \citenamefont {Wang}}]{Xu:2021uar}%
  \BibitemOpen
  \bibfield  {author} {\bibinfo {author} {\bibfnamefont {H.-j.}\ \bibnamefont {Xu}}, \bibinfo {author} {\bibfnamefont {W.}~\bibnamefont {Zhao}}, \bibinfo {author} {\bibfnamefont {H.}~\bibnamefont {Li}}, \bibinfo {author} {\bibfnamefont {Y.}~\bibnamefont {Zhou}}, \bibinfo {author} {\bibfnamefont {L.-W.}\ \bibnamefont {Chen}}, \ and\ \bibinfo {author} {\bibfnamefont {F.}~\bibnamefont {Wang}},\ }\href {\doibase 10.1103/PhysRevC.108.L011902} {\bibfield  {journal} {\bibinfo  {journal} {Phys. Rev. C}\ }\textbf {\bibinfo {volume} {108}},\ \bibinfo {pages} {L011902} (\bibinfo {year} {2023})},\ \Eprint {http://arxiv.org/abs/2111.14812} {arXiv:2111.14812 [nucl-th]} \BibitemShut {NoStop}%
\bibitem [{\citenamefont {Jia}(2022{\natexlab{b}})}]{Jia:2021qyu}%
  \BibitemOpen
  \bibfield  {author} {\bibinfo {author} {\bibfnamefont {J.}~\bibnamefont {Jia}},\ }\href {\doibase 10.1103/PhysRevC.105.044905} {\bibfield  {journal} {\bibinfo  {journal} {Phys. Rev. C}\ }\textbf {\bibinfo {volume} {105}},\ \bibinfo {pages} {044905} (\bibinfo {year} {2022}{\natexlab{b}})},\ \Eprint {http://arxiv.org/abs/2109.00604} {arXiv:2109.00604 [nucl-th]} \BibitemShut {NoStop}%
\bibitem [{\citenamefont {Liu}\ \emph {et~al.}(2022{\natexlab{a}})\citenamefont {Liu}, \citenamefont {Zhang}, \citenamefont {Zhou}, \citenamefont {Xu}, \citenamefont {Jia},\ and\ \citenamefont {Peng}}]{Liu:2022kvz}%
  \BibitemOpen
  \bibfield  {author} {\bibinfo {author} {\bibfnamefont {L.-M.}\ \bibnamefont {Liu}}, \bibinfo {author} {\bibfnamefont {C.-J.}\ \bibnamefont {Zhang}}, \bibinfo {author} {\bibfnamefont {J.}~\bibnamefont {Zhou}}, \bibinfo {author} {\bibfnamefont {J.}~\bibnamefont {Xu}}, \bibinfo {author} {\bibfnamefont {J.}~\bibnamefont {Jia}}, \ and\ \bibinfo {author} {\bibfnamefont {G.-X.}\ \bibnamefont {Peng}},\ }\href {\doibase 10.1016/j.physletb.2022.137441} {\bibfield  {journal} {\bibinfo  {journal} {Phys. Lett. B}\ }\textbf {\bibinfo {volume} {834}},\ \bibinfo {pages} {137441} (\bibinfo {year} {2022}{\natexlab{a}})},\ \Eprint {http://arxiv.org/abs/2203.09924} {arXiv:2203.09924 [nucl-th]} \BibitemShut {NoStop}%
\bibitem [{\citenamefont {Liu}\ \emph {et~al.}(2022{\natexlab{b}})\citenamefont {Liu}, \citenamefont {Zhang}, \citenamefont {Xu}, \citenamefont {Jia},\ and\ \citenamefont {Peng}}]{PhysRevC.106.034913}%
  \BibitemOpen
  \bibfield  {author} {\bibinfo {author} {\bibfnamefont {L.-M.}\ \bibnamefont {Liu}}, \bibinfo {author} {\bibfnamefont {C.-J.}\ \bibnamefont {Zhang}}, \bibinfo {author} {\bibfnamefont {J.}~\bibnamefont {Xu}}, \bibinfo {author} {\bibfnamefont {J.}~\bibnamefont {Jia}}, \ and\ \bibinfo {author} {\bibfnamefont {G.-X.}\ \bibnamefont {Peng}},\ }\href {\doibase 10.1103/PhysRevC.106.034913} {\bibfield  {journal} {\bibinfo  {journal} {Phys. Rev. C}\ }\textbf {\bibinfo {volume} {106}},\ \bibinfo {pages} {034913} (\bibinfo {year} {2022}{\natexlab{b}})}\BibitemShut {NoStop}%
\bibitem [{\citenamefont {Zhao}\ \emph {et~al.}(2023)\citenamefont {Zhao}, \citenamefont {Xu}, \citenamefont {Liu},\ and\ \citenamefont {Song}}]{Zhao:2022uhl}%
  \BibitemOpen
  \bibfield  {author} {\bibinfo {author} {\bibfnamefont {S.}~\bibnamefont {Zhao}}, \bibinfo {author} {\bibfnamefont {H.-j.}\ \bibnamefont {Xu}}, \bibinfo {author} {\bibfnamefont {Y.-X.}\ \bibnamefont {Liu}}, \ and\ \bibinfo {author} {\bibfnamefont {H.}~\bibnamefont {Song}},\ }\href {\doibase 10.1016/j.physletb.2023.137838} {\bibfield  {journal} {\bibinfo  {journal} {Phys. Lett. B}\ }\textbf {\bibinfo {volume} {839}},\ \bibinfo {pages} {137838} (\bibinfo {year} {2023})},\ \Eprint {http://arxiv.org/abs/2204.02387} {arXiv:2204.02387 [nucl-th]} \BibitemShut {NoStop}%
\bibitem [{\citenamefont {Jia}\ \emph {et~al.}(2023{\natexlab{a}})\citenamefont {Jia}, \citenamefont {Giacalone},\ and\ \citenamefont {Zhang}}]{Jia:2022qrq}%
  \BibitemOpen
  \bibfield  {author} {\bibinfo {author} {\bibfnamefont {J.}~\bibnamefont {Jia}}, \bibinfo {author} {\bibfnamefont {G.}~\bibnamefont {Giacalone}}, \ and\ \bibinfo {author} {\bibfnamefont {C.}~\bibnamefont {Zhang}},\ }\href {\doibase 10.1088/0256-307X/40/4/042501} {\bibfield  {journal} {\bibinfo  {journal} {Chin. Phys. Lett.}\ }\textbf {\bibinfo {volume} {40}},\ \bibinfo {pages} {042501} (\bibinfo {year} {2023}{\natexlab{a}})},\ \Eprint {http://arxiv.org/abs/2206.07184} {arXiv:2206.07184 [nucl-th]} \BibitemShut {NoStop}%
\bibitem [{\citenamefont {Giacalone}(2020)}]{Giacalone:2019pca}%
  \BibitemOpen
  \bibfield  {author} {\bibinfo {author} {\bibfnamefont {G.}~\bibnamefont {Giacalone}},\ }\href {\doibase 10.1103/PhysRevLett.124.202301} {\bibfield  {journal} {\bibinfo  {journal} {Phys. Rev. Lett.}\ }\textbf {\bibinfo {volume} {124}},\ \bibinfo {pages} {202301} (\bibinfo {year} {2020})},\ \Eprint {http://arxiv.org/abs/1910.04673} {arXiv:1910.04673 [nucl-th]} \BibitemShut {NoStop}%
\bibitem [{\citenamefont {Jia}\ \emph {et~al.}(2022)\citenamefont {Jia}, \citenamefont {Huang},\ and\ \citenamefont {Zhang}}]{Jia:2021wbq}%
  \BibitemOpen
  \bibfield  {author} {\bibinfo {author} {\bibfnamefont {J.}~\bibnamefont {Jia}}, \bibinfo {author} {\bibfnamefont {S.}~\bibnamefont {Huang}}, \ and\ \bibinfo {author} {\bibfnamefont {C.}~\bibnamefont {Zhang}},\ }\href {\doibase 10.1103/PhysRevC.105.014906} {\bibfield  {journal} {\bibinfo  {journal} {Phys. Rev. C}\ }\textbf {\bibinfo {volume} {105}},\ \bibinfo {pages} {014906} (\bibinfo {year} {2022})},\ \Eprint {http://arxiv.org/abs/2105.05713} {arXiv:2105.05713 [nucl-th]} \BibitemShut {NoStop}%
\bibitem [{\citenamefont {Bally}\ \emph {et~al.}(2022)\citenamefont {Bally}, \citenamefont {Bender}, \citenamefont {Giacalone},\ and\ \citenamefont {Som\`a}}]{Bally:2021qys}%
  \BibitemOpen
  \bibfield  {author} {\bibinfo {author} {\bibfnamefont {B.}~\bibnamefont {Bally}}, \bibinfo {author} {\bibfnamefont {M.}~\bibnamefont {Bender}}, \bibinfo {author} {\bibfnamefont {G.}~\bibnamefont {Giacalone}}, \ and\ \bibinfo {author} {\bibfnamefont {V.}~\bibnamefont {Som\`a}},\ }\href {\doibase 10.1103/PhysRevLett.128.082301} {\bibfield  {journal} {\bibinfo  {journal} {Phys. Rev. Lett.}\ }\textbf {\bibinfo {volume} {128}},\ \bibinfo {pages} {082301} (\bibinfo {year} {2022})},\ \Eprint {http://arxiv.org/abs/2108.09578} {arXiv:2108.09578 [nucl-th]} \BibitemShut {NoStop}%
\bibitem [{\citenamefont {Nijs}\ and\ \citenamefont {van~der Schee}(2023)}]{Nijs:2021kvn}%
  \BibitemOpen
  \bibfield  {author} {\bibinfo {author} {\bibfnamefont {G.}~\bibnamefont {Nijs}}\ and\ \bibinfo {author} {\bibfnamefont {W.}~\bibnamefont {van~der Schee}},\ }\href {\doibase 10.21468/SciPostPhys.15.2.041} {\bibfield  {journal} {\bibinfo  {journal} {SciPost Phys.}\ }\textbf {\bibinfo {volume} {15}},\ \bibinfo {pages} {041} (\bibinfo {year} {2023})},\ \Eprint {http://arxiv.org/abs/2112.13771} {arXiv:2112.13771 [nucl-th]} \BibitemShut {NoStop}%
\bibitem [{\citenamefont {Jia}\ and\ \citenamefont {Zhang}(2023)}]{Jia:2021oyt}%
  \BibitemOpen
  \bibfield  {author} {\bibinfo {author} {\bibfnamefont {J.}~\bibnamefont {Jia}}\ and\ \bibinfo {author} {\bibfnamefont {C.}~\bibnamefont {Zhang}},\ }\href {\doibase 10.1103/PhysRevC.107.L021901} {\bibfield  {journal} {\bibinfo  {journal} {Phys. Rev. C}\ }\textbf {\bibinfo {volume} {107}},\ \bibinfo {pages} {L021901} (\bibinfo {year} {2023})},\ \Eprint {http://arxiv.org/abs/2111.15559} {arXiv:2111.15559 [nucl-th]} \BibitemShut {NoStop}%
\bibitem [{\citenamefont {Chen}\ \emph {et~al.}(2024)\citenamefont {Chen} \emph {et~al.}}]{Chen:2024zwk}%
  \BibitemOpen
  \bibfield  {author} {\bibinfo {author} {\bibfnamefont {J.}~\bibnamefont {Chen}} \emph {et~al.},\ }\href {\doibase 10.1007/s41365-024-01591-2} {\bibfield  {journal} {\bibinfo  {journal} {Nucl. Sci. Tech.}\ }\textbf {\bibinfo {volume} {35}},\ \bibinfo {pages} {214} (\bibinfo {year} {2024})},\ \Eprint {http://arxiv.org/abs/2407.02935} {arXiv:2407.02935 [nucl-ex]} \BibitemShut {NoStop}%
\bibitem [{\citenamefont {Xu}\ \emph {et~al.}(2024)\citenamefont {Xu}, \citenamefont {Zhao},\ and\ \citenamefont {Wang}}]{Xu:2024bdh}%
  \BibitemOpen
  \bibfield  {author} {\bibinfo {author} {\bibfnamefont {H.-j.}\ \bibnamefont {Xu}}, \bibinfo {author} {\bibfnamefont {J.}~\bibnamefont {Zhao}}, \ and\ \bibinfo {author} {\bibfnamefont {F.}~\bibnamefont {Wang}},\ }\href {\doibase 10.1103/PhysRevLett.132.262301} {\bibfield  {journal} {\bibinfo  {journal} {Phys. Rev. Lett.}\ }\textbf {\bibinfo {volume} {132}},\ \bibinfo {pages} {262301} (\bibinfo {year} {2024})},\ \Eprint {http://arxiv.org/abs/2402.16550} {arXiv:2402.16550 [nucl-th]} \BibitemShut {NoStop}%
\bibitem [{\citenamefont {Wang}\ \emph {et~al.}(2024{\natexlab{a}})\citenamefont {Wang}, \citenamefont {Chen}, \citenamefont {Xu},\ and\ \citenamefont {Zhao}}]{Wang:2024vjf}%
  \BibitemOpen
  \bibfield  {author} {\bibinfo {author} {\bibfnamefont {Z.}~\bibnamefont {Wang}}, \bibinfo {author} {\bibfnamefont {J.}~\bibnamefont {Chen}}, \bibinfo {author} {\bibfnamefont {H.-j.}\ \bibnamefont {Xu}}, \ and\ \bibinfo {author} {\bibfnamefont {J.}~\bibnamefont {Zhao}},\ }\href {\doibase 10.1103/PhysRevC.110.034907} {\bibfield  {journal} {\bibinfo  {journal} {Phys. Rev. C}\ }\textbf {\bibinfo {volume} {110}},\ \bibinfo {pages} {034907} (\bibinfo {year} {2024}{\natexlab{a}})},\ \Eprint {http://arxiv.org/abs/2405.09329} {arXiv:2405.09329 [nucl-th]} \BibitemShut {NoStop}%
\bibitem [{\citenamefont {Zhao}\ \emph {et~al.}(2024{\natexlab{a}})\citenamefont {Zhao}, \citenamefont {Xu}, \citenamefont {Zhou}, \citenamefont {Liu},\ and\ \citenamefont {Song}}]{Zhao:2024lpc}%
  \BibitemOpen
  \bibfield  {author} {\bibinfo {author} {\bibfnamefont {S.}~\bibnamefont {Zhao}}, \bibinfo {author} {\bibfnamefont {H.-j.}\ \bibnamefont {Xu}}, \bibinfo {author} {\bibfnamefont {Y.}~\bibnamefont {Zhou}}, \bibinfo {author} {\bibfnamefont {Y.-X.}\ \bibnamefont {Liu}}, \ and\ \bibinfo {author} {\bibfnamefont {H.}~\bibnamefont {Song}},\ }\href {\doibase 10.1103/PhysRevLett.133.192301} {\bibfield  {journal} {\bibinfo  {journal} {Phys. Rev. Lett.}\ }\textbf {\bibinfo {volume} {133}},\ \bibinfo {pages} {192301} (\bibinfo {year} {2024}{\natexlab{a}})},\ \Eprint {http://arxiv.org/abs/2403.07441} {arXiv:2403.07441 [nucl-th]} \BibitemShut {NoStop}%
\bibitem [{\citenamefont {Xu}\ \emph {et~al.}(2022)\citenamefont {Xu}, \citenamefont {Li}, \citenamefont {Zhou}, \citenamefont {Wang}, \citenamefont {Zhao}, \citenamefont {Chen},\ and\ \citenamefont {Wang}}]{Xu:2021qjw}%
  \BibitemOpen
  \bibfield  {author} {\bibinfo {author} {\bibfnamefont {H.-j.}\ \bibnamefont {Xu}}, \bibinfo {author} {\bibfnamefont {H.}~\bibnamefont {Li}}, \bibinfo {author} {\bibfnamefont {Y.}~\bibnamefont {Zhou}}, \bibinfo {author} {\bibfnamefont {X.}~\bibnamefont {Wang}}, \bibinfo {author} {\bibfnamefont {J.}~\bibnamefont {Zhao}}, \bibinfo {author} {\bibfnamefont {L.-W.}\ \bibnamefont {Chen}}, \ and\ \bibinfo {author} {\bibfnamefont {F.}~\bibnamefont {Wang}},\ }\href {\doibase 10.1103/PhysRevC.105.L011901} {\bibfield  {journal} {\bibinfo  {journal} {Phys. Rev. C}\ }\textbf {\bibinfo {volume} {105}},\ \bibinfo {pages} {L011901} (\bibinfo {year} {2022})},\ \Eprint {http://arxiv.org/abs/2105.04052} {arXiv:2105.04052 [nucl-th]} \BibitemShut {NoStop}%
\bibitem [{\citenamefont {Zhao}\ and\ \citenamefont {Shi}(2023)}]{Zhao:2022mce}%
  \BibitemOpen
  \bibfield  {author} {\bibinfo {author} {\bibfnamefont {J.}~\bibnamefont {Zhao}}\ and\ \bibinfo {author} {\bibfnamefont {S.}~\bibnamefont {Shi}},\ }\href {\doibase 10.1140/epjc/s10052-023-11657-x} {\bibfield  {journal} {\bibinfo  {journal} {Eur. Phys. J. C}\ }\textbf {\bibinfo {volume} {83}},\ \bibinfo {pages} {511} (\bibinfo {year} {2023})},\ \Eprint {http://arxiv.org/abs/2211.01971} {arXiv:2211.01971 [hep-ph]} \BibitemShut {NoStop}%
\bibitem [{\citenamefont {Fortier}\ \emph {et~al.}(2025{\natexlab{a}})\citenamefont {Fortier}, \citenamefont {Jeon},\ and\ \citenamefont {Gale}}]{Fortier:2023xxy}%
  \BibitemOpen
  \bibfield  {author} {\bibinfo {author} {\bibfnamefont {N.~M.}\ \bibnamefont {Fortier}}, \bibinfo {author} {\bibfnamefont {S.}~\bibnamefont {Jeon}}, \ and\ \bibinfo {author} {\bibfnamefont {C.}~\bibnamefont {Gale}},\ }\href {\doibase 10.1103/PhysRevC.111.014901} {\bibfield  {journal} {\bibinfo  {journal} {Phys. Rev. C}\ }\textbf {\bibinfo {volume} {111}},\ \bibinfo {pages} {014901} (\bibinfo {year} {2025}{\natexlab{a}})},\ \Eprint {http://arxiv.org/abs/2308.09816} {arXiv:2308.09816 [nucl-th]} \BibitemShut {NoStop}%
\bibitem [{\citenamefont {Fortier}\ \emph {et~al.}(2025{\natexlab{b}})\citenamefont {Fortier}, \citenamefont {Jeon},\ and\ \citenamefont {Gale}}]{Fortier:2024yxs}%
  \BibitemOpen
  \bibfield  {author} {\bibinfo {author} {\bibfnamefont {N.~M.}\ \bibnamefont {Fortier}}, \bibinfo {author} {\bibfnamefont {S.}~\bibnamefont {Jeon}}, \ and\ \bibinfo {author} {\bibfnamefont {C.}~\bibnamefont {Gale}},\ }\href {\doibase 10.1103/PhysRevC.111.L011901} {\bibfield  {journal} {\bibinfo  {journal} {Phys. Rev. C}\ }\textbf {\bibinfo {volume} {111}},\ \bibinfo {pages} {L011901} (\bibinfo {year} {2025}{\natexlab{b}})},\ \Eprint {http://arxiv.org/abs/2405.17526} {arXiv:2405.17526 [nucl-th]} \BibitemShut {NoStop}%
\bibitem [{\citenamefont {Zhang}\ \emph {et~al.}(2024)\citenamefont {Zhang}, \citenamefont {Huang},\ and\ \citenamefont {Jia}}]{Zhang:2024bcb}%
  \BibitemOpen
  \bibfield  {author} {\bibinfo {author} {\bibfnamefont {C.}~\bibnamefont {Zhang}}, \bibinfo {author} {\bibfnamefont {S.}~\bibnamefont {Huang}}, \ and\ \bibinfo {author} {\bibfnamefont {J.}~\bibnamefont {Jia}},\ }\href@noop {} {\  (\bibinfo {year} {2024})},\ \Eprint {http://arxiv.org/abs/2405.08749} {arXiv:2405.08749 [nucl-th]} \BibitemShut {NoStop}%
\bibitem [{\citenamefont {Jia}\ \emph {et~al.}(2024{\natexlab{a}})\citenamefont {Jia}, \citenamefont {Huang}, \citenamefont {Zhang},\ and\ \citenamefont {Bhatta}}]{Jia:2024xvl}%
  \BibitemOpen
  \bibfield  {author} {\bibinfo {author} {\bibfnamefont {J.}~\bibnamefont {Jia}}, \bibinfo {author} {\bibfnamefont {S.}~\bibnamefont {Huang}}, \bibinfo {author} {\bibfnamefont {C.}~\bibnamefont {Zhang}}, \ and\ \bibinfo {author} {\bibfnamefont {S.}~\bibnamefont {Bhatta}},\ }\href@noop {} {\  (\bibinfo {year} {2024}{\natexlab{a}})},\ \Eprint {http://arxiv.org/abs/2408.15006} {arXiv:2408.15006 [nucl-th]} \BibitemShut {NoStop}%
\bibitem [{\citenamefont {Magdy}(2023{\natexlab{a}})}]{Magdy:2022cvt}%
  \BibitemOpen
  \bibfield  {author} {\bibinfo {author} {\bibfnamefont {N.}~\bibnamefont {Magdy}},\ }\href {\doibase 10.1140/epja/s10050-023-00982-0} {\bibfield  {journal} {\bibinfo  {journal} {Eur. Phys. J. A}\ }\textbf {\bibinfo {volume} {59}},\ \bibinfo {pages} {64} (\bibinfo {year} {2023}{\natexlab{a}})},\ \Eprint {http://arxiv.org/abs/2206.05332} {arXiv:2206.05332 [nucl-th]} \BibitemShut {NoStop}%
\bibitem [{\citenamefont {Ma}\ and\ \citenamefont {Zhang}(2022)}]{MaYG:2023}%
  \BibitemOpen
  \bibfield  {author} {\bibinfo {author} {\bibfnamefont {Y.-G.}\ \bibnamefont {Ma}}\ and\ \bibinfo {author} {\bibfnamefont {S.}~\bibnamefont {Zhang}},\ }\enquote {\bibinfo {title} {{Influence of Nuclear Structure in Relativistic Heavy-Ion Collisions}},}\ in\ \href {\doibase 10.1007/978-981-15-8818-1_5-1} {\emph {\bibinfo {booktitle} {{Handbook of Nuclear Physics}}}},\ \bibinfo {editor} {edited by\ \bibinfo {editor} {\bibfnamefont {I.}~\bibnamefont {Tanihata}}, \bibinfo {editor} {\bibfnamefont {H.}~\bibnamefont {Toki}}, \ and\ \bibinfo {editor} {\bibfnamefont {T.}~\bibnamefont {Kajino}}}\ (\bibinfo {year} {2022})\ pp.\ \bibinfo {pages} {1--30},\ \Eprint {http://arxiv.org/abs/2206.08218} {arXiv:2206.08218 [nucl-th]} \BibitemShut {NoStop}%
\bibitem [{\citenamefont {Xi}\ \emph {et~al.}(2025)\citenamefont {Xi}, \citenamefont {Chen}, \citenamefont {Ma}, \citenamefont {Ma},\ and\ \citenamefont {Wang}}]{XiBS:2025}%
  \BibitemOpen
  \bibfield  {author} {\bibinfo {author} {\bibfnamefont {B.~S.}\ \bibnamefont {Xi}}, \bibinfo {author} {\bibfnamefont {J.~H.}\ \bibnamefont {Chen}}, \bibinfo {author} {\bibfnamefont {L.}~\bibnamefont {Ma}}, \bibinfo {author} {\bibfnamefont {Y.~G.}\ \bibnamefont {Ma}}, \ and\ \bibinfo {author} {\bibfnamefont {T.~T.}\ \bibnamefont {Wang}},\ }\href {\doibase 10.1007/s41365-025-01826-w} {\bibfield  {journal} {\bibinfo  {journal} {Nucl. Sci. Tech.}\ }\textbf {\bibinfo {volume} {36}},\ \bibinfo {pages} {228} (\bibinfo {year} {2025})}\BibitemShut {NoStop}%
\bibitem [{\citenamefont {Giacalone}(2024)}]{GG:2024}%
  \BibitemOpen
  \bibfield  {author} {\bibinfo {author} {\bibfnamefont {G.}~\bibnamefont {Giacalone}},\ }\href {\doibase 10.1007/s41365-024-01582-3} {\bibfield  {journal} {\bibinfo  {journal} {Nucl. Sci. Tech.}\ }\textbf {\bibinfo {volume} {35}},\ \bibinfo {pages} {218} (\bibinfo {year} {2024})}\BibitemShut {NoStop}%
\bibitem [{\citenamefont {Schenke}(2024)}]{SB:2024}%
  \BibitemOpen
  \bibfield  {author} {\bibinfo {author} {\bibfnamefont {B.}~\bibnamefont {Schenke}},\ }\href {\doibase 10.1007/s41365-024-01509-y} {\bibfield  {journal} {\bibinfo  {journal} {Nucl. Sci. Tech.}\ }\textbf {\bibinfo {volume} {35}},\ \bibinfo {pages} {115} (\bibinfo {year} {2024})}\BibitemShut {NoStop}%
\bibitem [{\citenamefont {Schenke}(2025)}]{JJ:2025}%
  \BibitemOpen
  \bibfield  {author} {\bibinfo {author} {\bibfnamefont {B.}~\bibnamefont {Schenke}},\ }\href {\doibase 10.1007/s41365-025-01800-6} {\bibfield  {journal} {\bibinfo  {journal} {Nucl. Sci. Tech.}\ }\textbf {\bibinfo {volume} {36}},\ \bibinfo {pages} {207} (\bibinfo {year} {2025})}\BibitemShut {NoStop}%
\bibitem [{\citenamefont {Abdulhamid}\ \emph {et~al.}(2024)\citenamefont {Abdulhamid} \emph {et~al.}}]{STAR:2024wgy}%
  \BibitemOpen
  \bibfield  {author} {\bibinfo {author} {\bibfnamefont {M.~I.}\ \bibnamefont {Abdulhamid}} \emph {et~al.} (\bibinfo {collaboration} {STAR}),\ }\href {\doibase 10.1038/s41586-024-08097-2} {\bibfield  {journal} {\bibinfo  {journal} {Nature}\ }\textbf {\bibinfo {volume} {635}},\ \bibinfo {pages} {67} (\bibinfo {year} {2024})},\ \Eprint {http://arxiv.org/abs/2401.06625} {arXiv:2401.06625 [nucl-ex]} \BibitemShut {NoStop}%
\bibitem [{\citenamefont {Zhang}(2022)}]{Zhang:2022sgk}%
  \BibitemOpen
  \bibfield  {author} {\bibinfo {author} {\bibfnamefont {C.}~\bibnamefont {Zhang}} (\bibinfo {collaboration} {STAR}),\ }in\ \href@noop {} {\emph {\bibinfo {booktitle} {{10th International Conference on New Frontiers in Physics}}}}\ (\bibinfo {year} {2022})\ \Eprint {http://arxiv.org/abs/2203.13106} {arXiv:2203.13106 [nucl-ex]} \BibitemShut {NoStop}%
\bibitem [{\citenamefont {Aad}\ \emph {et~al.}(2023)\citenamefont {Aad} \emph {et~al.}}]{ATLAS:2022dov}%
  \BibitemOpen
  \bibfield  {author} {\bibinfo {author} {\bibfnamefont {G.}~\bibnamefont {Aad}} \emph {et~al.} (\bibinfo {collaboration} {ATLAS}),\ }\href {\doibase 10.1103/PhysRevC.107.054910} {\bibfield  {journal} {\bibinfo  {journal} {Phys. Rev. C}\ }\textbf {\bibinfo {volume} {107}},\ \bibinfo {pages} {054910} (\bibinfo {year} {2023})},\ \Eprint {http://arxiv.org/abs/2205.00039} {arXiv:2205.00039 [nucl-ex]} \BibitemShut {NoStop}%
\bibitem [{\citenamefont {Acharya}\ \emph {et~al.}(2022)\citenamefont {Acharya} \emph {et~al.}}]{ALICE:2021gxt}%
  \BibitemOpen
  \bibfield  {author} {\bibinfo {author} {\bibfnamefont {S.}~\bibnamefont {Acharya}} \emph {et~al.} (\bibinfo {collaboration} {ALICE}),\ }\href {\doibase 10.1016/j.physletb.2022.137393} {\bibfield  {journal} {\bibinfo  {journal} {Phys. Lett. B}\ }\textbf {\bibinfo {volume} {834}},\ \bibinfo {pages} {137393} (\bibinfo {year} {2022})},\ \Eprint {http://arxiv.org/abs/2111.06106} {arXiv:2111.06106 [nucl-ex]} \BibitemShut {NoStop}%
\bibitem [{\citenamefont {Acharya}\ \emph {et~al.}(2024)\citenamefont {Acharya} \emph {et~al.}}]{ALICE:2024nqd}%
  \BibitemOpen
  \bibfield  {author} {\bibinfo {author} {\bibfnamefont {S.}~\bibnamefont {Acharya}} \emph {et~al.} (\bibinfo {collaboration} {ALICE}),\ }\href@noop {} {\  (\bibinfo {year} {2024})},\ \Eprint {http://arxiv.org/abs/2409.04343} {arXiv:2409.04343 [nucl-ex]} \BibitemShut {NoStop}%
\bibitem [{\citenamefont {Zhang}\ \emph {et~al.}(2022{\natexlab{a}})\citenamefont {Zhang}, \citenamefont {Bhatta},\ and\ \citenamefont {Jia}}]{Zhang:2022fou}%
  \BibitemOpen
  \bibfield  {author} {\bibinfo {author} {\bibfnamefont {C.}~\bibnamefont {Zhang}}, \bibinfo {author} {\bibfnamefont {S.}~\bibnamefont {Bhatta}}, \ and\ \bibinfo {author} {\bibfnamefont {J.}~\bibnamefont {Jia}},\ }\href {\doibase 10.1103/PhysRevC.106.L031901} {\bibfield  {journal} {\bibinfo  {journal} {Phys. Rev. C}\ }\textbf {\bibinfo {volume} {106}},\ \bibinfo {pages} {L031901} (\bibinfo {year} {2022}{\natexlab{a}})},\ \Eprint {http://arxiv.org/abs/2206.01943} {arXiv:2206.01943 [nucl-th]} \BibitemShut {NoStop}%
\bibitem [{\citenamefont {Jia}\ \emph {et~al.}(2024{\natexlab{b}})\citenamefont {Jia} \emph {et~al.}}]{Jia:2022ozr}%
  \BibitemOpen
  \bibfield  {author} {\bibinfo {author} {\bibfnamefont {J.}~\bibnamefont {Jia}} \emph {et~al.},\ }\href {\doibase 10.1007/s41365-024-01589-w} {\bibfield  {journal} {\bibinfo  {journal} {Nucl. Sci. Tech.}\ }\textbf {\bibinfo {volume} {35}},\ \bibinfo {pages} {220} (\bibinfo {year} {2024}{\natexlab{b}})},\ \Eprint {http://arxiv.org/abs/2209.11042} {arXiv:2209.11042 [nucl-ex]} \BibitemShut {NoStop}%
\bibitem [{\citenamefont {Xu}(2023)}]{Xu:2022ikx}%
  \BibitemOpen
  \bibfield  {author} {\bibinfo {author} {\bibfnamefont {H.}~\bibnamefont {Xu}} (\bibinfo {collaboration} {STAR}),\ }\href {\doibase 10.5506/APhysPolBSupp.16.1-A30} {\bibfield  {journal} {\bibinfo  {journal} {Acta Phys. Polon. Supp.}\ }\textbf {\bibinfo {volume} {16}},\ \bibinfo {pages} {1} (\bibinfo {year} {2023})},\ \Eprint {http://arxiv.org/abs/2208.06149} {arXiv:2208.06149 [nucl-ex]} \BibitemShut {NoStop}%
\bibitem [{\citenamefont {Zhang}(2025)}]{Zhang:2024ake}%
  \BibitemOpen
  \bibfield  {author} {\bibinfo {author} {\bibfnamefont {C.}~\bibnamefont {Zhang}} (\bibinfo {collaboration} {STAR}),\ }\href {\doibase 10.1016/j.nuclphysa.2025.123106} {\bibfield  {journal} {\bibinfo  {journal} {Nucl. Phys. A}\ }\textbf {\bibinfo {volume} {1060}},\ \bibinfo {pages} {123106} (\bibinfo {year} {2025})},\ \Eprint {http://arxiv.org/abs/2409.09599} {arXiv:2409.09599 [nucl-ex]} \BibitemShut {NoStop}%
\bibitem [{\citenamefont {Lin}\ \emph {et~al.}(2005)\citenamefont {Lin}, \citenamefont {Ko}, \citenamefont {Li}, \citenamefont {Zhang},\ and\ \citenamefont {Pal}}]{Lin:2004en}%
  \BibitemOpen
  \bibfield  {author} {\bibinfo {author} {\bibfnamefont {Z.-W.}\ \bibnamefont {Lin}}, \bibinfo {author} {\bibfnamefont {C.~M.}\ \bibnamefont {Ko}}, \bibinfo {author} {\bibfnamefont {B.-A.}\ \bibnamefont {Li}}, \bibinfo {author} {\bibfnamefont {B.}~\bibnamefont {Zhang}}, \ and\ \bibinfo {author} {\bibfnamefont {S.}~\bibnamefont {Pal}},\ }\href {\doibase 10.1103/PhysRevC.72.064901} {\bibfield  {journal} {\bibinfo  {journal} {Phys. Rev. C}\ }\textbf {\bibinfo {volume} {72}},\ \bibinfo {pages} {064901} (\bibinfo {year} {2005})},\ \Eprint {http://arxiv.org/abs/nucl-th/0411110} {arXiv:nucl-th/0411110} \BibitemShut {NoStop}%
\bibitem [{\citenamefont {Wang}\ and\ \citenamefont {Gyulassy}(1991)}]{PhysRevD.44.3501}%
  \BibitemOpen
  \bibfield  {author} {\bibinfo {author} {\bibfnamefont {X.-N.}\ \bibnamefont {Wang}}\ and\ \bibinfo {author} {\bibfnamefont {M.}~\bibnamefont {Gyulassy}},\ }\href {\doibase 10.1103/PhysRevD.44.3501} {\bibfield  {journal} {\bibinfo  {journal} {Phys. Rev. D}\ }\textbf {\bibinfo {volume} {44}},\ \bibinfo {pages} {3501} (\bibinfo {year} {1991})}\BibitemShut {NoStop}%
\bibitem [{\citenamefont {Zhang}(1998)}]{Zhang:1997ej}%
  \BibitemOpen
  \bibfield  {author} {\bibinfo {author} {\bibfnamefont {B.}~\bibnamefont {Zhang}},\ }\href {\doibase 10.1016/S0010-4655(98)00010-1} {\bibfield  {journal} {\bibinfo  {journal} {Comput. Phys. Commun.}\ }\textbf {\bibinfo {volume} {109}},\ \bibinfo {pages} {193} (\bibinfo {year} {1998})},\ \Eprint {http://arxiv.org/abs/nucl-th/9709009} {arXiv:nucl-th/9709009} \BibitemShut {NoStop}%
\bibitem [{\citenamefont {He}\ and\ \citenamefont {Lin}(2017)}]{He:2017tla}%
  \BibitemOpen
  \bibfield  {author} {\bibinfo {author} {\bibfnamefont {Y.}~\bibnamefont {He}}\ and\ \bibinfo {author} {\bibfnamefont {Z.-W.}\ \bibnamefont {Lin}},\ }\href {\doibase 10.1103/PhysRevC.96.014910} {\bibfield  {journal} {\bibinfo  {journal} {Phys. Rev. C}\ }\textbf {\bibinfo {volume} {96}},\ \bibinfo {pages} {014910} (\bibinfo {year} {2017})},\ \Eprint {http://arxiv.org/abs/1703.02673} {arXiv:1703.02673 [nucl-th]} \BibitemShut {NoStop}%
\bibitem [{\citenamefont {Shao}\ \emph {et~al.}(2020)\citenamefont {Shao}, \citenamefont {Chen}, \citenamefont {Ko},\ and\ \citenamefont {Lin}}]{Shao:2020sqr}%
  \BibitemOpen
  \bibfield  {author} {\bibinfo {author} {\bibfnamefont {T.}~\bibnamefont {Shao}}, \bibinfo {author} {\bibfnamefont {J.}~\bibnamefont {Chen}}, \bibinfo {author} {\bibfnamefont {C.~M.}\ \bibnamefont {Ko}}, \ and\ \bibinfo {author} {\bibfnamefont {Z.-W.}\ \bibnamefont {Lin}},\ }\href {\doibase 10.1103/PhysRevC.102.014906} {\bibfield  {journal} {\bibinfo  {journal} {Phys. Rev. C}\ }\textbf {\bibinfo {volume} {102}},\ \bibinfo {pages} {014906} (\bibinfo {year} {2020})},\ \Eprint {http://arxiv.org/abs/2012.10037} {arXiv:2012.10037 [nucl-th]} \BibitemShut {NoStop}%
\bibitem [{\citenamefont {Li}\ and\ \citenamefont {Ko}(1995)}]{Li:1995pra}%
  \BibitemOpen
  \bibfield  {author} {\bibinfo {author} {\bibfnamefont {B.-A.}\ \bibnamefont {Li}}\ and\ \bibinfo {author} {\bibfnamefont {C.~M.}\ \bibnamefont {Ko}},\ }\href {\doibase 10.1103/PhysRevC.52.2037} {\bibfield  {journal} {\bibinfo  {journal} {Phys. Rev. C}\ }\textbf {\bibinfo {volume} {52}},\ \bibinfo {pages} {2037} (\bibinfo {year} {1995})},\ \Eprint {http://arxiv.org/abs/nucl-th/9505016} {arXiv:nucl-th/9505016} \BibitemShut {NoStop}%
\bibitem [{\citenamefont {M\"oller}\ \emph {et~al.}(2016)\citenamefont {M\"oller}, \citenamefont {Sierk}, \citenamefont {Ichikawa},\ and\ \citenamefont {Sagawa}}]{Moller:2015fba}%
  \BibitemOpen
  \bibfield  {author} {\bibinfo {author} {\bibfnamefont {P.}~\bibnamefont {M\"oller}}, \bibinfo {author} {\bibfnamefont {A.~J.}\ \bibnamefont {Sierk}}, \bibinfo {author} {\bibfnamefont {T.}~\bibnamefont {Ichikawa}}, \ and\ \bibinfo {author} {\bibfnamefont {H.}~\bibnamefont {Sagawa}},\ }\href {\doibase 10.1016/j.adt.2015.10.002} {\bibfield  {journal} {\bibinfo  {journal} {Atom. Data Nucl. Data Tabl.}\ }\textbf {\bibinfo {volume} {109-110}},\ \bibinfo {pages} {1} (\bibinfo {year} {2016})},\ \Eprint {http://arxiv.org/abs/1508.06294} {arXiv:1508.06294 [nucl-th]} \BibitemShut {NoStop}%
\bibitem [{\citenamefont {Zhang}\ and\ \citenamefont {Jia}(2022)}]{Zhang:2021kxj}%
  \BibitemOpen
  \bibfield  {author} {\bibinfo {author} {\bibfnamefont {C.}~\bibnamefont {Zhang}}\ and\ \bibinfo {author} {\bibfnamefont {J.}~\bibnamefont {Jia}},\ }\href {\doibase 10.1103/PhysRevLett.128.022301} {\bibfield  {journal} {\bibinfo  {journal} {Phys. Rev. Lett.}\ }\textbf {\bibinfo {volume} {128}},\ \bibinfo {pages} {022301} (\bibinfo {year} {2022})},\ \Eprint {http://arxiv.org/abs/2109.01631} {arXiv:2109.01631 [nucl-th]} \BibitemShut {NoStop}%
\bibitem [{\citenamefont {Jia}\ \emph {et~al.}(2023{\natexlab{b}})\citenamefont {Jia}, \citenamefont {Giacalone},\ and\ \citenamefont {Zhang}}]{Jia:2022qgl}%
  \BibitemOpen
  \bibfield  {author} {\bibinfo {author} {\bibfnamefont {J.}~\bibnamefont {Jia}}, \bibinfo {author} {\bibfnamefont {G.}~\bibnamefont {Giacalone}}, \ and\ \bibinfo {author} {\bibfnamefont {C.}~\bibnamefont {Zhang}},\ }\href {\doibase 10.1103/PhysRevLett.131.022301} {\bibfield  {journal} {\bibinfo  {journal} {Phys. Rev. Lett.}\ }\textbf {\bibinfo {volume} {131}},\ \bibinfo {pages} {022301} (\bibinfo {year} {2023}{\natexlab{b}})},\ \Eprint {http://arxiv.org/abs/2206.10449} {arXiv:2206.10449 [nucl-th]} \BibitemShut {NoStop}%
\bibitem [{\citenamefont {Lin}\ and\ \citenamefont {Ko}(2002)}]{Lin:2001zk}%
  \BibitemOpen
  \bibfield  {author} {\bibinfo {author} {\bibfnamefont {Z.-w.}\ \bibnamefont {Lin}}\ and\ \bibinfo {author} {\bibfnamefont {C.~M.}\ \bibnamefont {Ko}},\ }\href {\doibase 10.1103/PhysRevC.65.034904} {\bibfield  {journal} {\bibinfo  {journal} {Phys. Rev. C}\ }\textbf {\bibinfo {volume} {65}},\ \bibinfo {pages} {034904} (\bibinfo {year} {2002})},\ \Eprint {http://arxiv.org/abs/nucl-th/0108039} {arXiv:nucl-th/0108039} \BibitemShut {NoStop}%
\bibitem [{\citenamefont {Lin}\ and\ \citenamefont {Zheng}(2021)}]{Lin:2021mdn}%
  \BibitemOpen
  \bibfield  {author} {\bibinfo {author} {\bibfnamefont {Z.-W.}\ \bibnamefont {Lin}}\ and\ \bibinfo {author} {\bibfnamefont {L.}~\bibnamefont {Zheng}},\ }\href {\doibase 10.1007/s41365-021-00944-5} {\bibfield  {journal} {\bibinfo  {journal} {Nucl. Sci. Tech.}\ }\textbf {\bibinfo {volume} {32}},\ \bibinfo {pages} {113} (\bibinfo {year} {2021})},\ \Eprint {http://arxiv.org/abs/2110.02989} {arXiv:2110.02989 [nucl-th]} \BibitemShut {NoStop}%
\bibitem [{\citenamefont {Shen}\ \emph {et~al.}(2021)\citenamefont {Shen}, \citenamefont {Chen},\ and\ \citenamefont {Lin}}]{Shen:2021pds}%
  \BibitemOpen
  \bibfield  {author} {\bibinfo {author} {\bibfnamefont {D.}~\bibnamefont {Shen}}, \bibinfo {author} {\bibfnamefont {J.}~\bibnamefont {Chen}}, \ and\ \bibinfo {author} {\bibfnamefont {Z.-W.}\ \bibnamefont {Lin}},\ }\href {\doibase 10.1088/1674-1137/abe763} {\bibfield  {journal} {\bibinfo  {journal} {Chin. Phys. C}\ }\textbf {\bibinfo {volume} {45}},\ \bibinfo {pages} {054002} (\bibinfo {year} {2021})},\ \Eprint {http://arxiv.org/abs/2102.05266} {arXiv:2102.05266 [nucl-ex]} \BibitemShut {NoStop}%
\bibitem [{\citenamefont {Zhang}\ \emph {et~al.}(2022{\natexlab{b}})\citenamefont {Zhang}, \citenamefont {Chen}, \citenamefont {Li},\ and\ \citenamefont {Lin}}]{Zhang:2021ygs}%
  \BibitemOpen
  \bibfield  {author} {\bibinfo {author} {\bibfnamefont {L.}~\bibnamefont {Zhang}}, \bibinfo {author} {\bibfnamefont {J.}~\bibnamefont {Chen}}, \bibinfo {author} {\bibfnamefont {W.}~\bibnamefont {Li}}, \ and\ \bibinfo {author} {\bibfnamefont {Z.-W.}\ \bibnamefont {Lin}},\ }\href {\doibase 10.1016/j.physletb.2022.137063} {\bibfield  {journal} {\bibinfo  {journal} {Phys. Lett. B}\ }\textbf {\bibinfo {volume} {829}},\ \bibinfo {pages} {137063} (\bibinfo {year} {2022}{\natexlab{b}})},\ \Eprint {http://arxiv.org/abs/2112.14358} {arXiv:2112.14358 [hep-ph]} \BibitemShut {NoStop}%
\bibitem [{\citenamefont {Zhang}\ \emph {et~al.}(2023)\citenamefont {Zhang}, \citenamefont {Zheng}, \citenamefont {Shi},\ and\ \citenamefont {Lin}}]{Zhang:2022fum}%
  \BibitemOpen
  \bibfield  {author} {\bibinfo {author} {\bibfnamefont {C.}~\bibnamefont {Zhang}}, \bibinfo {author} {\bibfnamefont {L.}~\bibnamefont {Zheng}}, \bibinfo {author} {\bibfnamefont {S.}~\bibnamefont {Shi}}, \ and\ \bibinfo {author} {\bibfnamefont {Z.-W.}\ \bibnamefont {Lin}},\ }\href {\doibase 10.1016/j.physletb.2023.138219} {\bibfield  {journal} {\bibinfo  {journal} {Phys. Lett. B}\ }\textbf {\bibinfo {volume} {846}},\ \bibinfo {pages} {138219} (\bibinfo {year} {2023})},\ \Eprint {http://arxiv.org/abs/2210.07767} {arXiv:2210.07767 [nucl-th]} \BibitemShut {NoStop}%
\bibitem [{\citenamefont {Shao}\ \emph {et~al.}(2022)\citenamefont {Shao}, \citenamefont {Chen}, \citenamefont {Ma},\ and\ \citenamefont {Xu}}]{Shao:2022eyd}%
  \BibitemOpen
  \bibfield  {author} {\bibinfo {author} {\bibfnamefont {T.}~\bibnamefont {Shao}}, \bibinfo {author} {\bibfnamefont {J.}~\bibnamefont {Chen}}, \bibinfo {author} {\bibfnamefont {Y.-G.}\ \bibnamefont {Ma}}, \ and\ \bibinfo {author} {\bibfnamefont {Z.}~\bibnamefont {Xu}},\ }\href {\doibase 10.1103/PhysRevC.105.065801} {\bibfield  {journal} {\bibinfo  {journal} {Phys. Rev. C}\ }\textbf {\bibinfo {volume} {105}},\ \bibinfo {pages} {065801} (\bibinfo {year} {2022})},\ \Eprint {http://arxiv.org/abs/2205.13626} {arXiv:2205.13626 [hep-ph]} \BibitemShut {NoStop}%
\bibitem [{\citenamefont {Zhang}\ \emph {et~al.}(2025{\natexlab{a}})\citenamefont {Zhang}, \citenamefont {Chen},\ and\ \citenamefont {Zhang}}]{Zhang:2025yyd}%
  \BibitemOpen
  \bibfield  {author} {\bibinfo {author} {\bibfnamefont {L.}~\bibnamefont {Zhang}}, \bibinfo {author} {\bibfnamefont {J.}~\bibnamefont {Chen}}, \ and\ \bibinfo {author} {\bibfnamefont {C.}~\bibnamefont {Zhang}},\ }\href {\doibase 10.1103/PhysRevC.111.024911} {\bibfield  {journal} {\bibinfo  {journal} {Phys. Rev. C}\ }\textbf {\bibinfo {volume} {111}},\ \bibinfo {pages} {024911} (\bibinfo {year} {2025}{\natexlab{a}})},\ \Eprint {http://arxiv.org/abs/2501.08209} {arXiv:2501.08209 [nucl-th]} \BibitemShut {NoStop}%
\bibitem [{\citenamefont {Xu}\ and\ \citenamefont {Ko}(2011)}]{Xu:2011fi}%
  \BibitemOpen
  \bibfield  {author} {\bibinfo {author} {\bibfnamefont {J.}~\bibnamefont {Xu}}\ and\ \bibinfo {author} {\bibfnamefont {C.~M.}\ \bibnamefont {Ko}},\ }\href {\doibase 10.1103/PhysRevC.83.034904} {\bibfield  {journal} {\bibinfo  {journal} {Phys. Rev. C}\ }\textbf {\bibinfo {volume} {83}},\ \bibinfo {pages} {034904} (\bibinfo {year} {2011})},\ \Eprint {http://arxiv.org/abs/1101.2231} {arXiv:1101.2231 [nucl-th]} \BibitemShut {NoStop}%
\bibitem [{\citenamefont {Bilandzic}\ \emph {et~al.}(2011)\citenamefont {Bilandzic}, \citenamefont {Snellings},\ and\ \citenamefont {Voloshin}}]{PhysRevC.83.044913}%
  \BibitemOpen
  \bibfield  {author} {\bibinfo {author} {\bibfnamefont {A.}~\bibnamefont {Bilandzic}}, \bibinfo {author} {\bibfnamefont {R.}~\bibnamefont {Snellings}}, \ and\ \bibinfo {author} {\bibfnamefont {S.}~\bibnamefont {Voloshin}},\ }\href {\doibase 10.1103/PhysRevC.83.044913} {\bibfield  {journal} {\bibinfo  {journal} {Phys. Rev. C}\ }\textbf {\bibinfo {volume} {83}},\ \bibinfo {pages} {044913} (\bibinfo {year} {2011})}\BibitemShut {NoStop}%
\bibitem [{\citenamefont {Jia}\ \emph {et~al.}(2017)\citenamefont {Jia}, \citenamefont {Zhou},\ and\ \citenamefont {Trzupek}}]{Jia:2017hbm}%
  \BibitemOpen
  \bibfield  {author} {\bibinfo {author} {\bibfnamefont {J.}~\bibnamefont {Jia}}, \bibinfo {author} {\bibfnamefont {M.}~\bibnamefont {Zhou}}, \ and\ \bibinfo {author} {\bibfnamefont {A.}~\bibnamefont {Trzupek}},\ }\href {\doibase 10.1103/PhysRevC.96.034906} {\bibfield  {journal} {\bibinfo  {journal} {Phys. Rev. C}\ }\textbf {\bibinfo {volume} {96}},\ \bibinfo {pages} {034906} (\bibinfo {year} {2017})},\ \Eprint {http://arxiv.org/abs/1701.03830} {arXiv:1701.03830 [nucl-th]} \BibitemShut {NoStop}%
\bibitem [{\citenamefont {Lu}\ \emph {et~al.}(2023)\citenamefont {Lu}, \citenamefont {Zhao}, \citenamefont {Li}, \citenamefont {Jia},\ and\ \citenamefont {Zhou}}]{Lu:2023fqd}%
  \BibitemOpen
  \bibfield  {author} {\bibinfo {author} {\bibfnamefont {Z.}~\bibnamefont {Lu}}, \bibinfo {author} {\bibfnamefont {M.}~\bibnamefont {Zhao}}, \bibinfo {author} {\bibfnamefont {X.}~\bibnamefont {Li}}, \bibinfo {author} {\bibfnamefont {J.}~\bibnamefont {Jia}}, \ and\ \bibinfo {author} {\bibfnamefont {Y.}~\bibnamefont {Zhou}},\ }\href {\doibase 10.1140/epja/s10050-023-01194-2} {\bibfield  {journal} {\bibinfo  {journal} {Eur. Phys. J. A}\ }\textbf {\bibinfo {volume} {59}},\ \bibinfo {pages} {279} (\bibinfo {year} {2023})},\ \Eprint {http://arxiv.org/abs/2309.09663} {arXiv:2309.09663 [nucl-th]} \BibitemShut {NoStop}%
\bibitem [{\citenamefont {Magdy}(2023{\natexlab{b}})}]{Magdy:2022ize}%
  \BibitemOpen
  \bibfield  {author} {\bibinfo {author} {\bibfnamefont {N.}~\bibnamefont {Magdy}},\ }\href {\doibase 10.1103/PhysRevC.107.024905} {\bibfield  {journal} {\bibinfo  {journal} {Phys. Rev. C}\ }\textbf {\bibinfo {volume} {107}},\ \bibinfo {pages} {024905} (\bibinfo {year} {2023}{\natexlab{b}})},\ \Eprint {http://arxiv.org/abs/2210.14091} {arXiv:2210.14091 [nucl-th]} \BibitemShut {NoStop}%
\bibitem [{\citenamefont {Pei}\ \emph {et~al.}(2024)\citenamefont {Pei}, \citenamefont {Ma},\ and\ \citenamefont {Bzdak}}]{Pei:2024wsy}%
  \BibitemOpen
  \bibfield  {author} {\bibinfo {author} {\bibfnamefont {J.-L.}\ \bibnamefont {Pei}}, \bibinfo {author} {\bibfnamefont {G.-L.}\ \bibnamefont {Ma}}, \ and\ \bibinfo {author} {\bibfnamefont {A.}~\bibnamefont {Bzdak}},\ }\href {\doibase 10.1103/PhysRevC.110.024901} {\bibfield  {journal} {\bibinfo  {journal} {Phys. Rev. C}\ }\textbf {\bibinfo {volume} {110}},\ \bibinfo {pages} {024901} (\bibinfo {year} {2024})},\ \Eprint {http://arxiv.org/abs/2403.05782} {arXiv:2403.05782 [hep-ph]} \BibitemShut {NoStop}%
\bibitem [{\citenamefont {Acharya}\ \emph {et~al.}(2017)\citenamefont {Acharya} \emph {et~al.}}]{ALICE:2017fcd}%
  \BibitemOpen
  \bibfield  {author} {\bibinfo {author} {\bibfnamefont {S.}~\bibnamefont {Acharya}} \emph {et~al.} (\bibinfo {collaboration} {ALICE}),\ }\href {\doibase 10.1016/j.physletb.2017.07.060} {\bibfield  {journal} {\bibinfo  {journal} {Phys. Lett. B}\ }\textbf {\bibinfo {volume} {773}},\ \bibinfo {pages} {68} (\bibinfo {year} {2017})},\ \Eprint {http://arxiv.org/abs/1705.04377} {arXiv:1705.04377 [nucl-ex]} \BibitemShut {NoStop}%
\bibitem [{\citenamefont {{Chunjian Zhang, for the STAR Collaboration, WPCF 2022, Measurement of nuclear deformation in relativistic heavy-ion collisions at STAR}}()}]{chunjian}%
  \BibitemOpen
  \bibfield  {author} {\bibinfo {author} {\bibnamefont {{Chunjian Zhang, for the STAR Collaboration, WPCF 2022, Measurement of nuclear deformation in relativistic heavy-ion collisions at STAR}}},\ }\href@noop {} {}\BibitemShut {NoStop}%
\bibitem [{\citenamefont {B.~E.}\ \emph {et~al.}(2025)\citenamefont {B.~E.} \emph {et~al.}}]{STAR:2025vbp}%
  \BibitemOpen
  \bibfield  {author} {\bibinfo {author} {\bibfnamefont {A.}~\bibnamefont {B.~E.}} \emph {et~al.} (\bibinfo {collaboration} {STAR}),\ }\href {\doibase 10.1088/1361-6633/ae0fc3} {\bibfield  {journal} {\bibinfo  {journal} {Rept. Prog. Phys.}\ }\textbf {\bibinfo {volume} {88}},\ \bibinfo {pages} {108601} (\bibinfo {year} {2025})},\ \Eprint {http://arxiv.org/abs/2506.17785} {arXiv:2506.17785 [nucl-ex]} \BibitemShut {NoStop}%
\bibitem [{\citenamefont {Zhang}\ \emph {et~al.}(2025{\natexlab{b}})\citenamefont {Zhang}, \citenamefont {Jia}, \citenamefont {Chen}, \citenamefont {Shen},\ and\ \citenamefont {Liu}}]{Zhang:2025hvi}%
  \BibitemOpen
  \bibfield  {author} {\bibinfo {author} {\bibfnamefont {C.}~\bibnamefont {Zhang}}, \bibinfo {author} {\bibfnamefont {J.}~\bibnamefont {Jia}}, \bibinfo {author} {\bibfnamefont {J.}~\bibnamefont {Chen}}, \bibinfo {author} {\bibfnamefont {C.}~\bibnamefont {Shen}}, \ and\ \bibinfo {author} {\bibfnamefont {L.}~\bibnamefont {Liu}},\ }\href@noop {} {\  (\bibinfo {year} {2025}{\natexlab{b}})},\ \Eprint {http://arxiv.org/abs/2504.15245} {arXiv:2504.15245 [nucl-th]} \BibitemShut {NoStop}%
\bibitem [{\citenamefont {Abdallah}\ \emph {et~al.}(2022{\natexlab{b}})\citenamefont {Abdallah} \emph {et~al.}}]{STAR:2021mii}%
  \BibitemOpen
  \bibfield  {author} {\bibinfo {author} {\bibfnamefont {M.}~\bibnamefont {Abdallah}} \emph {et~al.} (\bibinfo {collaboration} {STAR}),\ }\href {\doibase 10.1103/PhysRevC.105.014901} {\bibfield  {journal} {\bibinfo  {journal} {Phys. Rev. C}\ }\textbf {\bibinfo {volume} {105}},\ \bibinfo {pages} {014901} (\bibinfo {year} {2022}{\natexlab{b}})},\ \Eprint {http://arxiv.org/abs/2109.00131} {arXiv:2109.00131 [nucl-ex]} \BibitemShut {NoStop}%
\bibitem [{\citenamefont {{Chunjian Zhang, for the STAR Collaboration, Quark Matter 2022, Observation of nuclear deformation in isobar collisions}}()}]{ChunjianQM}%
  \BibitemOpen
  \bibfield  {author} {\bibinfo {author} {\bibnamefont {{Chunjian Zhang, for the STAR Collaboration, Quark Matter 2022, Observation of nuclear deformation in isobar collisions}}},\ }\href@noop {} {\enquote {\bibinfo {title} {\url{https://indico.cern.ch/event/895086/contributions/4749420/attachments/2420958/4145197/QM2022_STAR_ChunjianZhang.pdf}},}\ }\BibitemShut {NoStop}%
\bibitem [{\citenamefont {Giacalone}\ \emph {et~al.}(2024)\citenamefont {Giacalone} \emph {et~al.}}]{Giacalone:2024luz}%
  \BibitemOpen
  \bibfield  {author} {\bibinfo {author} {\bibfnamefont {G.}~\bibnamefont {Giacalone}} \emph {et~al.},\ }\href@noop {} {\  (\bibinfo {year} {2024})},\ \Eprint {http://arxiv.org/abs/2402.05995} {arXiv:2402.05995 [nucl-th]} \BibitemShut {NoStop}%
\bibitem [{\citenamefont {Giacalone}\ \emph {et~al.}(2025)\citenamefont {Giacalone} \emph {et~al.}}]{Giacalone:2024ixe}%
  \BibitemOpen
  \bibfield  {author} {\bibinfo {author} {\bibfnamefont {G.}~\bibnamefont {Giacalone}} \emph {et~al.},\ }\href {\doibase 10.1103/PhysRevLett.134.082301} {\bibfield  {journal} {\bibinfo  {journal} {Phys. Rev. Lett.}\ }\textbf {\bibinfo {volume} {134}},\ \bibinfo {pages} {082301} (\bibinfo {year} {2025})},\ \Eprint {http://arxiv.org/abs/2405.20210} {arXiv:2405.20210 [nucl-th]} \BibitemShut {NoStop}%
\bibitem [{\citenamefont {Zhang}\ \emph {et~al.}(2025{\natexlab{c}})\citenamefont {Zhang}, \citenamefont {Chen}, \citenamefont {Giacalone}, \citenamefont {Huang}, \citenamefont {Jia},\ and\ \citenamefont {Ma}}]{Zhang:2024vkh}%
  \BibitemOpen
  \bibfield  {author} {\bibinfo {author} {\bibfnamefont {C.}~\bibnamefont {Zhang}}, \bibinfo {author} {\bibfnamefont {J.}~\bibnamefont {Chen}}, \bibinfo {author} {\bibfnamefont {G.}~\bibnamefont {Giacalone}}, \bibinfo {author} {\bibfnamefont {S.}~\bibnamefont {Huang}}, \bibinfo {author} {\bibfnamefont {J.}~\bibnamefont {Jia}}, \ and\ \bibinfo {author} {\bibfnamefont {Y.-G.}\ \bibnamefont {Ma}},\ }\href {\doibase 10.1016/j.physletb.2025.139322} {\bibfield  {journal} {\bibinfo  {journal} {Phys. Lett. B}\ }\textbf {\bibinfo {volume} {862}},\ \bibinfo {pages} {139322} (\bibinfo {year} {2025}{\natexlab{c}})},\ \Eprint {http://arxiv.org/abs/2404.08385} {arXiv:2404.08385 [nucl-th]} \BibitemShut {NoStop}%
\bibitem [{\citenamefont {Zhao}\ \emph {et~al.}(2024{\natexlab{b}})\citenamefont {Zhao}, \citenamefont {Ma}, \citenamefont {Zhou}, \citenamefont {Lin},\ and\ \citenamefont {Zhang}}]{Zhao:2024feh}%
  \BibitemOpen
  \bibfield  {author} {\bibinfo {author} {\bibfnamefont {X.-L.}\ \bibnamefont {Zhao}}, \bibinfo {author} {\bibfnamefont {G.-L.}\ \bibnamefont {Ma}}, \bibinfo {author} {\bibfnamefont {Y.}~\bibnamefont {Zhou}}, \bibinfo {author} {\bibfnamefont {Z.-W.}\ \bibnamefont {Lin}}, \ and\ \bibinfo {author} {\bibfnamefont {C.}~\bibnamefont {Zhang}},\ }\href@noop {} {\  (\bibinfo {year} {2024}{\natexlab{b}})},\ \Eprint {http://arxiv.org/abs/2404.09780} {arXiv:2404.09780 [nucl-th]} \BibitemShut {NoStop}%
\bibitem [{\citenamefont {Wang}\ \emph {et~al.}(2024{\natexlab{b}})\citenamefont {Wang}, \citenamefont {Zhao}, \citenamefont {Cao}, \citenamefont {Xu},\ and\ \citenamefont {Song}}]{YuanyuanWang:2024sgp}%
  \BibitemOpen
  \bibfield  {author} {\bibinfo {author} {\bibfnamefont {Y.}~\bibnamefont {Wang}}, \bibinfo {author} {\bibfnamefont {S.}~\bibnamefont {Zhao}}, \bibinfo {author} {\bibfnamefont {B.}~\bibnamefont {Cao}}, \bibinfo {author} {\bibfnamefont {H.-j.}\ \bibnamefont {Xu}}, \ and\ \bibinfo {author} {\bibfnamefont {H.}~\bibnamefont {Song}},\ }\href {\doibase 10.1103/PhysRevC.109.L051904} {\bibfield  {journal} {\bibinfo  {journal} {Phys. Rev. C}\ }\textbf {\bibinfo {volume} {109}},\ \bibinfo {pages} {L051904} (\bibinfo {year} {2024}{\natexlab{b}})},\ \Eprint {http://arxiv.org/abs/2401.15723} {arXiv:2401.15723 [nucl-th]} \BibitemShut {NoStop}%
\bibitem [{\citenamefont {Wang}\ \emph {et~al.}(2024{\natexlab{c}})\citenamefont {Wang}, \citenamefont {Li}, \citenamefont {Liu}, \citenamefont {Xu},\ and\ \citenamefont {Ren}}]{Wang:2024ulq}%
  \BibitemOpen
  \bibfield  {author} {\bibinfo {author} {\bibfnamefont {H.-C.}\ \bibnamefont {Wang}}, \bibinfo {author} {\bibfnamefont {S.-J.}\ \bibnamefont {Li}}, \bibinfo {author} {\bibfnamefont {L.-M.}\ \bibnamefont {Liu}}, \bibinfo {author} {\bibfnamefont {J.}~\bibnamefont {Xu}}, \ and\ \bibinfo {author} {\bibfnamefont {Z.-Z.}\ \bibnamefont {Ren}},\ }\href {\doibase 10.1103/PhysRevC.110.034909} {\bibfield  {journal} {\bibinfo  {journal} {Phys. Rev. C}\ }\textbf {\bibinfo {volume} {110}},\ \bibinfo {pages} {034909} (\bibinfo {year} {2024}{\natexlab{c}})},\ \Eprint {http://arxiv.org/abs/2409.02452} {arXiv:2409.02452 [nucl-th]} \BibitemShut {NoStop}%
\bibitem [{\citenamefont {Liu}\ \emph {et~al.}(2024)\citenamefont {Liu}, \citenamefont {Li}, \citenamefont {Wang}, \citenamefont {Xu}, \citenamefont {Ren},\ and\ \citenamefont {Huang}}]{Liu:2023gun}%
  \BibitemOpen
  \bibfield  {author} {\bibinfo {author} {\bibfnamefont {L.-M.}\ \bibnamefont {Liu}}, \bibinfo {author} {\bibfnamefont {S.-J.}\ \bibnamefont {Li}}, \bibinfo {author} {\bibfnamefont {Z.}~\bibnamefont {Wang}}, \bibinfo {author} {\bibfnamefont {J.}~\bibnamefont {Xu}}, \bibinfo {author} {\bibfnamefont {Z.-Z.}\ \bibnamefont {Ren}}, \ and\ \bibinfo {author} {\bibfnamefont {X.-G.}\ \bibnamefont {Huang}},\ }\href {\doibase 10.1016/j.physletb.2024.138724} {\bibfield  {journal} {\bibinfo  {journal} {Phys. Lett. B}\ }\textbf {\bibinfo {volume} {854}},\ \bibinfo {pages} {138724} (\bibinfo {year} {2024})},\ \Eprint {http://arxiv.org/abs/2312.13572} {arXiv:2312.13572 [nucl-th]} \BibitemShut {NoStop}%
\bibitem [{\citenamefont {Lu}\ \emph {et~al.}(2025)\citenamefont {Lu}, \citenamefont {Zhao}, \citenamefont {Nielsen}, \citenamefont {Li},\ and\ \citenamefont {Zhou}}]{Lu:2025cni}%
  \BibitemOpen
  \bibfield  {author} {\bibinfo {author} {\bibfnamefont {Z.}~\bibnamefont {Lu}}, \bibinfo {author} {\bibfnamefont {M.}~\bibnamefont {Zhao}}, \bibinfo {author} {\bibfnamefont {E.~G.~D.}\ \bibnamefont {Nielsen}}, \bibinfo {author} {\bibfnamefont {X.}~\bibnamefont {Li}}, \ and\ \bibinfo {author} {\bibfnamefont {Y.}~\bibnamefont {Zhou}},\ }\href {\doibase 10.1016/j.physletb.2025.139698} {\bibfield  {journal} {\bibinfo  {journal} {Phys. Lett. B}\ }\textbf {\bibinfo {volume} {868}},\ \bibinfo {pages} {139698} (\bibinfo {year} {2025})},\ \Eprint {http://arxiv.org/abs/2501.14852} {arXiv:2501.14852 [nucl-th]} \BibitemShut {NoStop}%
\end{thebibliography}%
\end{document}